\documentstyle[12pt]{article}
\renewcommand{\theequation}{\thesection.\arabic{equation}}

\def\ut#1{$\underline{\smash{\vphantom{y}\hbox{#1}}}$}
\def\bra#1{\mathopen{\langle#1\,|}}
\def\braa#1{\mathopen{\langle#1\,||}}
\def\ket#1{\mathclose{|\,#1\rangle}}
\def\kett#1{\mathclose{||\,#1\rangle}}
\def\dq3{d^3 q}

\def\ninj#1#2#3#4#5#6#7#8#9{\left\{\begin{array}{ccc}
                                  #1&#2&#3\\
                                  #4&#5&#6\\
                                  #7&#8&#9
                             \end{array}\right\}}

\def\sla#1{\rlap\slash #1}

\newcommand{\be}{\begin{eqnarray}}
\newcommand{\ee}{\end{eqnarray}}
\newcommand{\BE} {\begin{equation}}
\newcommand{\EE} {\end{equation}}
\newcommand{\BEAnn}{\begin{eqnarray*}}
\newcommand{\EEAnn}{\end{eqnarray*}}
\def\lrarr#1{\vbox{\ialign{##\crcr
    $\leftrightarrow$\crcr
    \noalign{\kern 1pt\nointerlineskip}
    $\hfil\displaystyle{#1}\hfil$\crcr}}}
\def\dsp{\displaystyle}
\def\ov#1#2{\dsp #1\over\dsp #2}
\def\eps{\epsilon}
\newcommand{\head}
{\setlength{\unitlength}{0.8 mm}
\raisebox{-3 mm}
{\begin{picture}(20,10) (-10,-5)
\put(-10,5){\line(2,-1){20}}
\put(2.0,2.2){\makebox(0,0){$T$}}
\put(-2.0,-2.2){\makebox(0,0){$P$}}
\end{picture}}}
\def\sfil{\phantom{\begin{tabular}{c}
           $K_{ss}(2)$ $K_{sv}^{(+)}(21)$ $K_{sv}^{(-)}(22)$ $K_{vv}(211)$
           \\
           $K_{vv}(202)$ $K_{vv}(220)$ $K_{vv}(222)$
     \end{tabular}}}

\def\lsim{\mathrel{\rlap{\lower4pt\hbox{\hskip1pt$\sim$}}
    \raise1pt\hbox{$<$}}}         %less than or approx. symbol
\renewcommand{\thefootnote}{*}
\title{\bf Time Reversal Invariance in the $\beta$--Decays
  of A = 8 Nuclei}
\author{S. Ying\footnotemark ~ and E. M. Henley\\
{\it Department of Physics, FM-15, University of Washington}\\
\it Seattle, WA 98195}
\date{}
\begin{document}
\maketitle
\footnotetext{Present address:
Physics Department, Fudan University, Shanghai 200433, P. R. China.}
\renewcommand{\thefootnote}{\thesection.\arabic{footnote}}
\setcounter{footnote}{0}
\vspace{4 mm}
\begin{abstract}
We examine time reversal invariance in the $\beta$--decays of $^8$B(2$^+$) and
$^8$Li(2$^+$) to $^8$Be (2$^+$) in detail, with particular attention to final
state interactions of the two $\alpha$ particles from the decay of $^8$Be, and
 of
the electron (positron) with the daughter nucleus.  An R--matrix formulation is
used, with the initial state described by a shell model.  The R--matrix
parameters are obtained by fitting the $\alpha-\alpha$ scattering phase shifts
and allowed $\beta$--decay rates. Because the nuclear final state interaction
effects on the time reversal test come from second forbidden $\beta$--decays,
they are small and can be minimized by
a suitable choice of kinematic conditions.  The $e^\pm$--nucleus Coulomb
interaction induced T--violation effects depend on the first forbidden
operators.
\end{abstract}
%\input{sections}
%
%sections.tex
%
\setcounter{equation}{0}
\setcounter{footnote}{0}
\renewcommand{\thefootnote}{\thesection.\arabic{footnote}}
\section{Introduction}
Tests of time reversal invariance (TRI) remain very important because it is
only in the $K^0$--$\bar K^0$ system that CP (or T) invariance has been
found not to hold to about $2\times 10^{-3}$ \cite{epspexp}. Until TRI
violations (TRIV) have
been found in other systems, it is difficult to distinguish between various
theoretical proposals [2-7] that have
been put forward to explain the experimental results obtained in the decays
of neutral kaons.

Nuclear tests of TRI that have been carried out include detailed
balance \cite{AlMg}
($a_T\leq 5\times 10^{-4}$), reciprocity \cite{LAMPF} ($a_T\leq 10^{-2}$),
polarized hyperon decay \cite{Lambdatoppion,TRIbook}, nuclear $\gamma$--decay
\cite{CalTech} and nuclear $\beta$--decay \cite{Dcoef1}
($a_T\leq 10^{-3}$) experiments. The most accurate test of TRI, combined with
 parity
non-conservation is the lack of neutron and atomic electric dipole
moments \cite{epspexp,Foston}.
For a review of experiments and theory, we refer to
Ref. \cite{TRIbook,HenleySumm}.

Tests of TRI in $\beta$--decay are made difficult
by the necessity of measuring a triple correlation or the polarization of
the emitted electron.  The triple correlation $\langle\vec J\rangle\cdot
\vec p_e\times\vec p_\nu$ requires a polarized parent state and the
measurement of the recoil nucleus.  The mass eight system is amenable to a
somewhat simpler test, because the daughter of the decays from $^8$Li or
$^8$B, namely $^8$Be$^*$, is unstable and breaks up into two charged particles
($^4$He) which can be detected more easily than the recoil.  However, a
question arises immediately: Does the final state interaction amongst the
final $e^\pm$ and the two $\alpha$ particles emitted in the decay of
$^8$Be$^*$ spoil the TRI test? In this work we investigate this question in
detail.  We find that true tests of TRI which depend on measurements of one
(or both) recoil $\alpha$ particles are only spoiled by competing second
order forbidden $\beta$--decays and by the $e^\pm$--nucleus Coulomb
interactions. By a proper choice
of kinematics, even this small spoiling can be minimized.  The primary
decay is from a 2$^+$ state of $^8$Li or $^8$B to the lowest 2$^+$ state of
$^8$Be$^*$.  ``Pseudo--TRI violation" (PTRIV) for the strong interaction
effects, which mimics
TRIV through the introduction of
phases \cite{Jacobhen} occur primarily through the interference of the main
2$^+$ $\to$ 2$^+$ matrix elements with those of the $2^+\to 0^+$ and $2^+\to
{4^+}$ ($J=0^+$ and 4$^+$) of $^8$Be; the
PTRIV effects for the $e^\pm$--nucleus Coulomb interaction depend on the
Coulomb phase shifts and the $2^+ \to 2^+$ transition operators.

A complete first--principle theoretical study is immensely complex.  We treat
the decay as a two--step process (see Fig.~1)
\be
^8{\rm Li} (^8{\rm B}) &\to& ^8{\rm Be}^* + e^-(e^+)+\bar\nu(\nu)\nonumber\\
&\to& \alpha+\alpha + e^-(e^+)+\bar\nu(\nu),\nonumber
\ee
and use an R--matrix formalism in which the initial state is described by a
shell model wave function and the $^8$Be$^*$ resonant states are saturated by
shell
model states.  The parameters of the R--matrix are constrained by physical
considerations and are adjusted to reproduce $\alpha$--$\alpha$ scattering
phase shifts in the relevant energy region.  With these parameters, it is
found that the unpolarized $\beta$--decay spectrum is reproduced with two
additional parameters.  The TRIV and PTRIV correlation observables are
studied in  this framework.

In the next section, we examine the TRI tests in the mass 8 system. In
Section 3, we develop the R--matrix approximation to be used. In Section 4, we
study the PTRIV due to hadronic interactions; in
Section 5 we study that due to the $e^\pm$--nucleus Coulomb interactions, and
finally
in Section 6 we summarize our results.

\setcounter{equation}{0}
\setcounter{footnote}{0}
\renewcommand{\thefootnote}{\thesection.\arabic{footnote}}
\section{Time reversal invariance tests in the $\beta$--decay processes
of the A=8 System}

The questions related to tests of TRI in the weak and
electromagnetic processes have been studied by many authors.
We restrict ourselves here to the mass 8 system, for which an
experimental study has been proposed \cite{Adelberger}. The initial 2$^+$
$^8$Li or $^8$B ground states decay by $\beta$--emission
to the (resonant) states of $^8$Be$^*$ with $J^P = 0^+,2^+,4^+,\ldots$, which
break up into two $\alpha$ particles. The graphical
representation, together with definitions of some kinematic
variables are given in Fig. 1.
The final $2^+$ resonant states
of $^8$Be$^*$ are our main interest. Those states with $J^P=0^+$ and 4$^+$ can
interfere with the $2^+$ state to produce PTRIV
signals due to final (2$\alpha$) state interactions.
These states and interactions should be taken into account in the study
of TRIV effects. The resonant states of $^8$Be$^*$ are
treated in section 3. In the present
section, emphasis is given to presenting the analytic results and
the shell model computation of the hadronic weak transition amplitudes.

\subsection{The decay rates for the unpolarized case}

      The differential decay rate for the $\beta$--decay of an unpolarized
nucleus can be expressed as
\be
 dW &=& 2\pi\delta(E_i-E_f-\epsilon_e-\epsilon_\nu)
Tr \left [H_W^\dagger \Theta H_W\right ] d\xi,\label{dW1}
\ee
where $d\xi$ is an infinitesimal phase space element of the final state,
$\Theta$ is the final state projection operator,
$H_W$ is the weak Hamiltonian responsible for the weak charged current nuclear
reactions
and the trace is over the initial and final state (spin + linear momentum)
subspaces.
We write
\be
Tr\left [H_W^\dagger\Theta H_W\right ] &=& {1\over 2}
G_F^2 cos^2\theta_c
W^{\mu\nu} L_{\mu\nu}.\label{TrHH}
\ee
When the leptonic polarizations are not detected, $W^{\mu\nu}$ and $L^{\mu\nu}$
are
\be
W^{\mu\nu} &=& Tr\left [J^{\nu \dagger}(-{\bf q})\Theta J^\mu(-{\bf q})
\right ],\label{Wmn}\\
L^{\mu\nu} &=& Tr\left [l^{\nu\dagger}({\bf q})
l^\mu({\bf q})\right ],\label{Lmn}
\ee
where $J^\mu({\bf q})$ is the Fourier transform of the hadronic weak
current operator $J^\mu({\bf x},t=0)$,
$l_\mu({\bf q})$ is Fourier transform of the matrix element of the
leptonic weak current operators with ${\bf q } = - {\bf k}_e - {\bf
k}_\nu$.\footnote{The Coulomb interactions between the charged leptons and
nuclei that are ignored here will be investigated in Section 5.}
Eq.  \ref{dW1} can be written as
\be
dW &=& 2\pi\delta(E_i-E_f-\epsilon_e-\epsilon_\nu)\nonumber\\ &&
{1\over 2} G_F^2 cos^2\theta_c\left [
W_{ss} L_{ss} + W^{(+)i}_{sv} L^{(+)i}_{sv} +
W^{(-)i}_{sv} L^{(-)i}_{sv} + W^{ij}_{vv} L^{ij}_{vv}\right ]d\xi, \nonumber\\
&&\label{dW2}
\ee
where $i,j=1\ldots 3$ and irreducible tensor components of the
leptonic tensors $L_{ss},\ldots,L_{vv}$ are given in the following
subsection. $W_{ss}$ and $W_{vv}$ contain  contributions of the time
(labeled by ``$s$'') and space (labeled by ``$v$'') components of the
hadronic weak current operators. $W^{(\pm)}_{sv}$ contains
contributions of the interference terms between the time and space
components of the same current operator, where ``+'' (``--'') denotes the
symmetric (anti--symmetric) combination with respect to the time and space
indices of the corresponding operator.

\subsection{Kinematic functions and hadronic response functions}

In the decay of $^8$B and $^8$Li, the daughter $^8$Be$^*$ breaks up into two
$\alpha$ particles.
The resonant state of the two $\alpha$ particles is of the following form
\be
\ket{-;\hat{\bf k}}_{2\alpha} &=& 4\pi\sum_{J_f = 0^+,2^+,4^+}
\sum_{M_f = -J_f}^{J_f} i^{J_f} e^{-i\delta_{J_f}} Y_{J_fM_f}^*(\hat{\bf k})
\ket{J_fM_f},\label{BeFinal}
\ee
where ``--'' denotes an incoming boundary condition,
$\hat{\bf k}$ is a unit vector in the direction of
the relative momentum of the two $\alpha$
particles, $\delta_{J_f}$
is the strong and electromagnetic
phase shift of the $J_f$ partial wave and $Y_{JM}$ is a
spherical harmonic.

Using the trace formula given in  \cite{thesis}, various terms in
Eq.  \ref{dW2} can be expressed in terms of a set of irreducible
hadronic response functions, namely,
\be
W_{ss} L_{ss} &=& (4\pi)^2\sum_{J_f{J'}_f} i^{{J'}_f-J_f} e^{i(\delta_{J_f} -
\delta_{{J'}_f})} \sum_{\sigma}
C_\sigma^{J_f{J'}_f} R_{ss}(\sigma) K_{ss}(\sigma),
\label{W1L1}\\
W_{sv}^{(\pm)i}
L_{sv}^{(\pm)i} &=&(4\pi)^2\sum_{J_f{J'}_f} i^{{J'}_f-J_f} e^{i(\delta_{J_f} -
\delta_{{J'}_f})}
\sum_{\sigma\rho} C_\sigma^{J_f{J'}_f}
R^{(\pm)}_{sv}(\sigma\rho) K^{(\pm)}_{sv}(\sigma\rho), \nonumber\\
&&\label{W2L2}\\
W_{vv}^{ij} L_{vv}^{ij} &=&
(4\pi)^2\sum_{J_f{J'}_f} i^{{J'}_f-J_f} e^{i(\delta_{J_f} -
\delta_{{J'}_f})}
\sum_{\sigma\rho\tau}C_\sigma^{J_f{J'}_f}
R_{vv}(\sigma\rho\tau) K_{vv}(\sigma\rho\tau), \nonumber\\
&&\label{W4L4}
\ee
where
\be
C_\sigma^{J_f{J'}_f} &=& \sqrt{4\pi}
{\braa{{J'}_f}Y_\sigma\kett{J_f}\over \tilde J_i\tilde\sigma}\label{Csigma}
\ee
with $\tilde x \equiv \sqrt{2 x + 1}$.
The kinematic functions are
\be
K_{ss}(\sigma) &=& 4\pi \left [ Y_\sigma(\hat{\bf k})\otimes Y_\sigma(\hat{\bf
q})\right ]_0 L_{ss},\label{Kss1}\\
K^{(\pm)}_{sv}(\sigma\rho) &=& 4\pi \left [ Y_\sigma(\hat{\bf k})\otimes
[Y_\rho(\hat{\bf q})\otimes L^{(\pm)}_{sv}]_\sigma\right ]_0,\label{Ksv+1}\\
K_{vv}(\sigma\rho\tau) &=& 4\pi \left [ Y_\sigma(\hat{\bf k})\otimes
[Y_\rho(\hat{\bf q})\otimes L_{vv}^\tau]_\sigma\right ]_0,\label{Kvv1}
\ee
where $[\phi_{l_1}\otimes\phi_{l_2}]_{jm}$
denotes the Clebsch--Gordan coupling of $\phi_{l_1m_1}$ and $\phi_{l_2m_2}$ to
total angular momentum j and magnetic quantum number m. The total number of
different kinematic functions are 34; 22 of them are T--even and 12 of them
are T--odd.

It can be shown that
the leading terms of the hadronic response functions
$R_{sv}^{(\pm)}(\sigma\rho)$
and $R_{vv}(\sigma\rho\tau)$ are of order $O[(R\kappa)^\rho]$ (it is $O[(R
\kappa)^\sigma]$ for $R_{ss}(\sigma)$),
where $R$ is a typical radius of the parent/daughter nucleus;
thus at low momentum transfer, terms
with large $\rho$ can be neglected since
$R\kappa<<1$. If one keeps terms up to order
$O[\kappa^2R^2]$, the possible kinematic functions with $\sigma\le 2$
together with their properties
under parity and time reversal transformations are listed in Table
 \ref{KineFunc}. The T--odd correlation observables interested in this
paper are P--odd. There are four of them which are given in the lower right
box of Table  \ref{KineFunc}.

The leptonic tensors in case of not observing the polarization of the
charged leptons are found to be
\be
L_{ss} &=& 8 \left [\epsilon_e\epsilon_\nu + {\bf k}_e\cdot{\bf k}_\nu
\right ],
\label{Lssup}\\
L_{sv}^{(+)} &=& 8 \left [\epsilon_e {\bf k}_\nu + \epsilon_\nu {\bf k}_e
\right ],
\label{Lsvpup}\\
L_{sv}^{(-)} &=& \mp 8 i {\bf k}_e\times {\bf k}_\nu,
\label{Lsvmup}\\
L_{vv}^0 &=& -{8\over\sqrt{3}} \left [3\epsilon_e\epsilon_\nu -{\bf k}_e\cdot
{\bf k}_\nu\right ],
\label{Lvv0up}\\
L_{vv}^1 &=& \pm 8\sqrt{2} \left
[\epsilon_\nu {\bf k}_e - \epsilon_e {\bf k}_\nu \right ],
\label{Lvv1up}\\
L_{vv}^2 &=& 8 \left [{\bf k}_e{\bf k}_\nu + {\bf k}_\nu {\bf k}_e - {2\over 3}
                {\bf k}_e\cdot {\bf k}_\nu
               \right ].\label{Lvv2up}
\ee

The hadronic dynamic response functions
$R_{ss},\ldots,R_{vv}$ can be expressed
in terms of the reduced matrix elements of the multipole
operators \cite{Walecka}
$C_J,\ldots,M^5_J$ of the hadronic charged weak currents. The analytic form
of these relations can be found in Appendix A. The T--odd response functions,
which are of interest in this work, have the following explicit expressions.

\noindent
\underline{Contribution from the transition to the $J^P=2^+$ states:}
\be
R_{sv}^{(+)}(22) &=& i{2\over 5} ImC_0M_2^{5*} + i{1\over 5}\sqrt{3\over 2}
                     ReC_2E_1^{5*},
\label{Rto1}\\
R_{sv}^{(-)}(21) &=& -i{2\over 5}\sqrt{3\over 5} ImC_0E_2^* -i{2\over
                     5}\sqrt{2\over 5}ImC_0L_2^*-i{2\over 25}\sqrt{7} ImL_1^5
                     C_1^{5*}\nonumber\\
                  &&  -i{3\over 25}\sqrt{7\over 2}ImE_1^5C_1^{5*} +
                      i{2\over 5}\sqrt{2\over 5}ImL_0C_2^*,
\label{Rto2}\\
R_{vv}(221)      &=& i{1\over 10}\sqrt{21\over 5} ImL_1^5E_1^{5*} - i{\sqrt{2}
                       \over 5}ImL_0E_2^* + i{2\over 5}\sqrt{3\over 35}
                       ImL^5_3E_1^{5*}\nonumber\\
                 &&  -i{4\over 5}\sqrt{2\over 35} ImL_1^5E_3^{5*}
                     +i{1\over 10} ReL_1^5M_2^{5*},
\label{Rto3}\\
R_{vv}(212)      &=& -i{\sqrt{21}\over 50} Im L_1^5 M_1^*
                     -i{2\over 25}\sqrt{21\over 2} Im E_1^5M_1^*
                    +i{1\over 5}\sqrt{2\over 5} Im L_0M_2^{5*}\nonumber\\
                  &&  +i{1\over 5}\sqrt{2\over 5} ReE_1^5 E_2^*
                    +i{1\over 10}\sqrt{3\over 5}ReE_1^5 L_2^*
                    -i{1\over 10\sqrt{5}} ReL_1^5E_2^*.
\label{Rto4}
\ee
Here, symbols with a superscript ``5'' indicate that they are
contributions of the hadronic axial vector current operator; others
are contributions of the hadronic vector current operator. $C_J$ or
$C^5_J$ are reduced matrix elements of the multipole operators of the
weak charge density operators. $L_J\ldots M_J^5$ are matrix elements of
the space components of the weak current operators. $L_J$ or $L_J^5$
are reduced matrix
elements of the multipole operators of the longitudinal part of the
weak current operators. $E_J$ or $E^5_J$ are reduced
matrix elements of the electric part of the transverse components of the
weak current multipole operators and $M_J$ or $M_J^5$ are the reduced
matrix elements of the magnetic part of the transverse components of
the same operators.

\noindent
\underline{Contribution from the interference terms between $J^P=2^+$ and
$J^P=0^+$ states:}
\be
R_{sv}^{(+)}(22) &=& {1\over 5} C_0M_2^{5*} -{1\over 5}\sqrt{3\over 2}
                      E_1^5 C_2^* + {\sqrt{2}\over 10}C_1E_2^*,
\label{Rto5}\\
R_{sv}^{(-)}(21) &=& -{1\over 5}\sqrt{3\over 5} C_0 E_2^* + {1\over 5}
                     \sqrt{2\over 5} L_0 C_2^* -{1\over 5}\sqrt{2\over 5}
                    C_0L_2^*,
\label{Rto6}\\
R_{vv}(221)    &=& -{\sqrt{2}\over 10} L_0 E_2^* - {1\over 10}  L_1^5
                     M_2^{5*},
\label{Rto7}\\
R_{vv}(212)     &=& {1\over 10}\sqrt{2\over 5}  L_0M_2^{5*} - {1\over 5}
                     \sqrt{2\over 5}  E_1^5 E_2^*
                     -{1\over 10}\sqrt{3\over 5} E_1^5 L_2^*\nonumber\\
                &&   +{\sqrt{5}\over 50}  L_1^5 E_2^*,
\label{Rto8}
\ee
where those reduced matrix elements that are complex conjugated correspond
to transitions to the final $0^+$ state and all the rest correspond
to the transition to the final $2^+$ state.

\noindent
\underline{Contribution from the interference terms between $J^P=2^+$ and
$J^P=4^+$ states:}
\be
R_{sv}^{(+)}(22) &=& {1\over 5}  C_0 M_2^{5*} + {1\over 5}\sqrt{2\over 5}
                   E_1^5 C_2^*-{\sqrt{2}\over 15} C_1E_2^*,
\label{Rto9}\\
R_{sv}^{(-)}(21) &=& -{1\over 5}\sqrt{3\over 5}  C_0 E_2^* + {1\over 5}
                 \sqrt{2\over 5}  L_0 C_2^*-{1\over 5}\sqrt{2\over 5}
                 C_0L_2^*,
\label{Rto10}\\
R_{vv}(221)   &=& -{\sqrt{2}\over 10}  L_0 E_2^* +
               {1\over 15}  L_1^5 M_2^{5*}
\label{Rto11}\\
R_{vv}(212)   &=& {1\over 10}\sqrt{2\over 5} L_0 M_2^{5*} + {2\over 15}
                 \sqrt{2\over 5}  E_1^5 E_2^* + {1\over 25}\sqrt{5\over 3}
                 E_1^5 L_2^*\nonumber\\
               && -{\sqrt{5}\over 75}  L_1^5 E_2^*,
\label{Rto12}
\ee
where those reduced matrix elements that are complex conjugated correspond
to transitions to the final $4^+$ state and all the rest correspond
to the transition to the final $2^+$ state.

The T--odd observables corresponding
to these interference response functions depend on their imaginary parts.

The T--even response functions can also be written in the same expanded  form
as those given above for the T--odd ones. They will not be presented here;
their expressions can be deduced by using the compact expressions given
in Appendix A.

\subsection{The differential decay date in terms of the weak
transition amplitudes}

The kinematic functions can be written in Cartesian form.
The differential decay rate Eq.  \ref{dW2} in the $2^+$ resonant region is
\be
dW &=& dW_{22}^{(e)}
       + dW_{22}^{(o)} + dW_{20}^{(o)} + dW_{24}^{(o)}+\ldots,\label{dW3}
\ee
where $dW_{22}^{(e)}$ is a contribution to the
differential decay rate even under time reversal
and $dW_{JJ'}^{(o)}$ are contributions to differential decay rates odd under
time reversal.
The subindices denote the contributing angular
momentum ($J$) of the partial waves of the final two $\alpha$ particles.
The relatively small
terms that originate from the interference between the $2^+$
and the $0^+$ and $4^+$ final states
are included for the T--odd observables only since they are important
contributions to the PTRIV that originate from the final state
interaction. The contributions from pure $0^+$ and $4^+$ states to the
TRIV signal are negligible in the energy region where the lowest
$2^+$ state resonates; they are not included in the differential decay rate.

When $m_e$ is neglected,
the T--even differential decay rate
has the following generic form
\be
dW_{22}^{(e)} &=& dW_0\left\{ \left [
             R_1^{(0)} + R_2^{(0)}\hat{\bf k}_e\cdot\hat{\bf k}_\nu
             +R_3^{(0)}\left (\left(
             \hat{\bf k}_e\cdot\hat{\bf k}_\nu\right)^2-{1\over 3}
             \right )\right ]\right. \nonumber\\
            &&+  T_2(\hat{\bf k}):\left [\left (
             R^{(2)}_1\hat{\bf k}_e\hat{\bf k}_e
            + R^{(2)}_2\hat{\bf k}_e\hat{\bf k}_\nu
            + R^{(2)}_3\hat{\bf k}_\nu\hat{\bf k}_\nu\right )\right.\nonumber\\
            && +\left.\left.\left
               (R_4^{(2)}\hat{\bf k}_e\hat{\bf k}_e
            + R^{(2)}_5\hat{\bf k}_e\hat{\bf k}_\nu
            + R^{(2)}_6\hat{\bf k}_\nu\hat{\bf k}_\nu\right )\hat{\bf k}_e
            \cdot \hat{\bf k}_\nu\right ]\right \},\label{dWe}
\ee
where the leading decay rate $dW_0$ is
\be
dW_0 &=& (2\pi)^3\delta(E_i-E_f-\epsilon_e-\epsilon_\nu){16\over 5}\epsilon_e
        \epsilon_\nu G_F^2
        cos^2\theta_cF(Z,\eps_e) |g_A|^2 |A_2|^2 d\xi,\nonumber\\
&&\label{dW0}
\ee
and the non--relativistic one body
Gamow--Teller matrix element $A_2$ is defined as
\be
A_2&=&\braa{2^+}{\cal  Y}_{20}\cdot{\sigma}\kett{2^+}\label{A2}
\ee
with ${\cal  Y}_{Jlm} = \left [Y_l\otimes\hat\epsilon \right]_{Jm}$,
$\hat{\epsilon}^\dagger\cdot\hat{\epsilon}=1$ and $F(Z,\eps_e)$ the Coulomb
function for the charged leptons. In Eq.  \ref{dW0}, the tensorial
contraction between the second
rank tensor $T_2(\hat {\bf k})$ and a pair of vectors ${\bf AB}$ is defined
as
\be
   T_2(\hat{\bf k}):{\bf AB} &\equiv& T_2(\hat{\bf
k})_{ij}A_iB_j,\label{T2AB}\\
   T_2(\hat{\bf k})_{ij} &=& \hat k_i\hat k_j - {1\over 3}\delta_{ij}.
\label{T2ij}
\ee
The neglect of $m_e$ is justifiable since the energy released in
the transitions ($\sim$ 3--10 MeV) is much larger than $m_e$.

Three T--odd differential rates are
\be
dW_{22}^{(0)} &=& dW_0 T_2(\hat {\bf k}):\left [ R_7^{(2)}\hat {\bf k}_e
\hat{\bf k}_e\times\hat{\bf k}_\nu + R_8^{(2)} \hat {\bf k}_\nu
\hat{\bf k}_e\times\hat {\bf k}_\nu
\right ],\label{TodddW22}\\
dW_{20}^{(0)} &=& dW_0 T_2(\hat {\bf k}):\left [
R_9^{(2)}\hat {\bf k}_e
\hat{\bf k}_e\times\hat{\bf k}_\nu + R_{10}^{(2)} \hat {\bf k}_\nu
\hat{\bf k}_e\times\hat {\bf k}_\nu
\right ],\label{TodddW20}\\
dW_{24}^{(0)} &=& dW_0 T_2(\hat {\bf k}):\left
[ R_{11}^{(2)}\hat{\bf k}_e
\hat{\bf k}_e\times\hat {\bf k}_\nu + R_{12}^{(2)} \hat{\bf k}_\nu
\hat{\bf k}_e\times\hat {\bf k}_\nu
\right ].\label{TodddW42}
\ee
where we also have neglected $m_e$.

The Cartesian response functions $R_i^{(0)}$ $(i=1,\ldots,3)$ and
$R_i^{(2)}$ $(i=1,\ldots,12)$
are related to those in the spherical harmonics
basis \cite{thesis}, namely $R_{ss},\ldots,R_{vv}$. They are given in
Appendix B. At low momentum transfers, we expand these
momentum dependent multipole operators of the hadronic
charged weak currents in terms of their static multipoles, up to
order of $O(\kappa/M)$ with $M$ the mass of a nucleon, and/or
$O(R^2\kappa^2)$ with $R\sim 1-2$ fm the size of a
typical nucleus. As a first
step, we take the non--relativistic one body approximation
for the hadronic charged weak current operators by retaining only the
leading $1\over M$ term in an expansion. The justification of such a step can
be understood following the discussions of Ref. [20].  The results are well
known \cite{Walecka}. They are reproduced in Appendix C for completeness.

The result of the expansion of the hadronic charged weak currents
operator in the non--relativistic one body
approximation
and the definition of the related static multipole operators
are given in Tables  \ref{MultOpExpand} and  \ref{StaticMop}.
The symbol $\braa{}\ldots\kett{}$ represents the reduced matrix elements.

It is useful to introduce the following quantities
\be
\eta_i &=& {A_i\over A_2},\label{eta}
\ee
with $A_i$ defined in Table  \ref{StaticMop}, and
\be
   f_1 &=& -{F_1\over g_A},\label{lowf1}\\
   f_M &=& -\left ({G^V_M\over g_A}+\sqrt{2\over 3} f_1\eta_4\right ),
\label{lowfM}\\
  f_c^5&=& 1 + {2\over\sqrt 3}\eta_3,\label{lowfc5}\\
  f_T  &=& -{g_T\over g_A},\label{lowfT}
\ee
where the isospin indices ``$(\pm)$'' for the single nucleon form factors
are suppressed.

The Cartesian form of the T--odd response functions $R_{7}^{(2)},\ldots,
R_{12}^{(2)}$ are
\be
R_7^{(2)}&=&{\eps_e\over 2M}\left (\mp Imf_c^5 + Imf_M - Imf_T\right )
\nonumber\\
          &&  + {\eps_e\Delta\over 2M^2}\left ( {1\over 7}\sqrt{7\over 15}
             Imf_1\eta_6 \pm {\sqrt{2}\over 10} Im\eta_8
             \pm {1\over 30}\sqrt{10\over 7} Im\eta_9
              \pm {4\over 15}{1\over\sqrt{7}} Im\eta_{10}\right )
\nonumber\\
          && + {\eps_e^2\over 2M^2} \left ( -{2\over 7}\sqrt{7\over 15}
            Imf_1\eta_6\mp{\sqrt{2}\over 5}Im\eta_8\mp{1\over
15}\sqrt{10\over7}
            Im\eta_9 \pm {8\over 15\sqrt{7}} Im\eta_{10}\right ),\nonumber\\
 &&\label{Rt1}\\
R_8^{(2)}&=&-{\Delta-\eps_e\over 2M}\left (\pm Imf_c^5 + Imf_M + Imf_T\right )
\nonumber\\
          && + {\Delta^2\over M^2}\left ({1\over 14}\sqrt{7\over 15}
            Imf_1\eta_6\pm {\sqrt{2}\over 10}Im\eta_8
           \mp {1\over 30}\sqrt{10\over
            7}Im\eta_9\pm {4\over 15\sqrt{7}} Im\eta_{10}\right )
\nonumber\\
        && -{\Delta\eps_e\over M^2}\left ({5 \over 14} \sqrt{7\over 15} Imf_1
              \eta_6
           \pm {3\sqrt{2} \over 10}Im\eta_8\mp {1 \over 30} \sqrt{10 \over 7}
            Im\eta_9
           \pm{4\over 5\sqrt{7}} Im\eta_{10}\right )
\nonumber\\
        && - {\eps_e^2\over M^2}\left (
           -{2\over 7}\sqrt{7\over 15} Imf_1\eta_6\mp{\sqrt{2}\over 5}
            Im\eta_8\mp {1\over 15}\sqrt{10\over 7}
            Im\eta_9\mp {8\over 15\sqrt{7}}
            Im\eta_{10}\right ),\nonumber\\
       &&\label{Rt2}\\
R_9^{(2)}&=&-{\eps_e\Delta\over M^2}sin(\delta_2-\delta_0)
\left ( {1\over 2}\sqrt{2\over 3} Re f_1\eta_6(2^+,0^+)\pm {1\over 15}
Re\eta_9(2^+,0^+)\right )
\nonumber \\
  && + {2\eps_e^2\over M^2} sin(\delta_2-\delta_0)
\left ({2\over 5}\sqrt{2\over 3}Ref_1\eta_6(2^+,0^+)
\pm{1\over 15}Re\eta_9(2^+,0^+)\right ),\label{Rt3}\\
R_{10}^{(2)}&=&-{\Delta^2\over M^2}sin(\delta_2-\delta_0)
                \left ( {3\over 10}\sqrt{2\over 3}
                 Ref_1\eta_6(2^+,0^+)\pm{1\over 15}Re\eta_9(2^+,0^+) \right )
\nonumber\\
            && + {\Delta\eps_e\over M^2} sin(\delta_2-\delta_0)
                \left ( {11\over 10}\sqrt{2\over 3}
                Ref_1\eta_6(2^+,0^+)\pm {1\over 5}Re\eta_9(2^+,0^+)\right )
\nonumber\\
            && -{2\eps_e^2\over M^2} sin(\delta_2-\delta_0)
                \left ( {2\over 5}\sqrt{2\over 3}
                Ref_1\eta_6(2^+,0^+)\pm {1\over 15}Re\eta_9(2^+,0^+)\right ),
\label{Rt4}\\
R_{11}^{(2)}&=&{\eps_e\Delta\over M^2}
               sin(\delta_2-\delta_4)\left ({2\over 5}\sqrt{3\over 7}Ref_1
               \eta_6(2^+,4^+)\pm {2\over 15}\sqrt{2\over 7}
                Re\eta_9(2^+,4^+)\right )
\nonumber\\
             &&-{2\eps_e^2\over M^2}sin(\delta_2-\delta_4)
                \left ({4\over 15}\sqrt{3\over 7}Ref_1\eta_6(2^+,4^+)
                \pm {2\over 15}\sqrt{2\over 7} Re\eta_9(2^+,4^+)\right ),
\nonumber \\ &&\label{Rt5}\\
R_{12}^{(2)}&=&{\Delta^2\over M^2}sin(\delta_2-\delta_4)\left
                ({2\over 15}\sqrt{3\over 7}Ref_1\eta_6(2^+,4^+)
                \pm {2\over 15}\sqrt{2\over 7}Re\eta_9(2^+,4^+)\right )
\nonumber\\
             &&-{\Delta\eps_e\over M^2} sin(\delta_2-\delta_4)\left
               ({2\over 3}\sqrt{3\over 7}
                Ref_1\eta_6(2^+,4^+)\pm {2\over 15}\sqrt{2\over 7}
                Re\eta_9(2^+,4^+)\right )
\nonumber\\
             &&+{2\eps_e^2\over M^2} sin(\delta_2-\delta_4)\left
              ({4\over 15}\sqrt{3\over 7}
               Ref_1\eta_6(2^+,4^+)\pm {2\over 15}\sqrt{2\over 7}
               Re\eta_9(2^+,4^+)\right ),\nonumber \\
&& \label{Rt6}
\ee
where
\be
\eta_6(2^+,J^+) &=& M^2{\dsp\sum_{i=1}^A\braa{J^+}\tau^{(\pm)}
                 r^2Y_2\kett{2^+}\over
                 \dsp\sum_{i=1}^A\braa{2^+}\tau^{(\pm)}{\cal Y}_{10}
                    \cdot\sigma\kett{2^+}},\label{eta620}\\
\eta_9(2^+,J^+) &=& M^2 {\dsp\sum_{i=1}^A\braa{J^+}\tau^{(\pm)}
                    r^2{\cal Y}_{22}\cdot\sigma\kett{2^+}
                    \over\dsp\sum_{i=1}^A\braa{2^+}\tau^{(\pm)}
                    {\cal Y}_{10}\cdot\sigma\kett{2^+}},
                    \label{eta920}\\
\ee
with $J^P=0^+,4^+$.
Terms in Eqs. \ref{Rt3}--\ref{Rt6} that are proportional to $cos(\ldots)$
have been dropped, since they are much smaller than
those terms proportional
to $sin(\ldots)$ due to the fact they are multiplied by T--odd hadronic
response functions which are expected to be much smaller than the
corresponding T--even components even if the Hamiltonian of the system is
not invariant under time reversal. The Fermi matrix element $A_1$ is
assumed to be zero.

\subsection{Shell model calculation of the matrix elements of the
        static multipole operators in the A=8 system}

The intermediate
states of the $\beta$--decay processes are $^8$Be$^*$ resonances. Here, we
shall
consider a shell model computation of the reduced matrix elements of
various static multipole operators that are used
in the next section where the decay of these shell model states are
discussed and where a comparison with experiments is made.

We shall restrict ourselves to a $1p$ shell model space. Using the effective
interaction developed by Cohen and Kurath \cite{CohenK}(2BME 8--16), the shell
model Hamiltonian can be diagonalized \cite{Haxton1}.
The $1p$ configuration mixing
shell model calculation with the Cohen--Kurath potential (2BME 8--16)
is used to calculate the $J^P=2^+$ levels
 of $^8$Be$^*$ (see Fig. 2);
these are stable shell model states that lie
at E=3.55(0), 13.37(0), 16.19(1), 17.45(0),...(the numbers
in parentheses are the isospin for the corresponding states and
the unit is in MeV).
These theoretically determined energy levels lie reasonably
close to the experimental resonant state energies given in Fig. 3.
The shell model
state at E=3.55 corresponds to the E=3.04 state and that
at E=16.19 corresponds to the E=16.63 state in Fig. 2.
The shell model
state at E=13.37 possesses similar one particle parentage coefficient
in its wave function \cite{SixteenState} as the shell model state located at
E=16.92, so although it is relatively far from the state at E=16.92,
it should be identified with that state.
The E=17.45 state corresponds to some higher excitation state in
the $J^P=2^+$ state series; it is not considered in this paper.

The doubly reduced matrix element of an one body transition
operator has the following form \cite{Walecka,Haxton2}
\be
<J_1;T_1\vdots\vdots O^{(1)}_{J;T} \vdots\vdots J_2; T_2>
&=& \sum_{aa'}\psi_{J;T}(a'a) <a'\vdots\vdots O^{(1)}_{J;T}
\vdots\vdots a>,\label{OneBOp}
\ee
where $a,a'$ enumerate single nucleon shell model states. In the $1p$
shell model subspace, $a=j=(1/2,3/2)$,
with $j$ the single nucleon total angular momentum.
The density matrix $\psi_{J;T}(a',a)$ is defined in Ref. \cite{Haxton2};
it depends on the shell model Hamiltonian in the $1p$ shell
subspace. The reduced matrix elements of various static multipole operators
defined
in Table  \ref{StaticMop} are
\be
A_i &=& \pm {1\over\sqrt{3}}Tr\psi^T {\bf A}_i,\label{AiFinal}
\ee
where ``T'' denotes transpose and $\psi$ is the corresponding density matrix.

By using harmonic oscillator radial wave functions \cite{Haxton2},
the one nucleon matrix elements corresponding to those defined in
Table  \ref{StaticMop} are
\def\factor{{<I_T>\over\sqrt{4\pi}}}
\be
{\bf A}_1&=&\factor \pmatrix{\sqrt{2}&0\cr 0 & 2\cr},\label{A1}\\&&\\
{\bf A}_2&=&\factor \pmatrix{-\sqrt{2\over 3} & 4\over\sqrt{3}\cr
                       -4\over\sqrt{3} & 2\sqrt{5\over 3}\cr},\label{A1A2}\\
&&\\
{\bf A}_3
&=&\factor \pmatrix{1\over\sqrt{2}&-5\cr -1 & -\sqrt{5}\cr},\label{Aa2}\\
&&\\
{\bf A}_4
&=&\factor \pmatrix{-2 & \sqrt{2}\cr-\sqrt{2}&-\sqrt{10}\cr},\label{A3A4}\\
&&\\
{\bf A}_5
&=&\factor \pmatrix{5\over\sqrt 2 & 0\cr 0 & 5\cr}(Mb)^2,\label{A5}\\
&&\\
{\bf A}_6
&=&\factor \pmatrix{0 & 5 \cr -5 & -5\cr}(Mb)^2,\label{A5A6}\\
&&\\
{\bf A}_7
&=&\factor \pmatrix{-{5\over 2}\sqrt{2\over 3} & 10\over\sqrt{3}\cr
                       -{10\over\sqrt{3}} & 5\sqrt{5\over 3}\cr}
                 (Mb)^2,\label{A7}\\
&&\\
{\bf A}_8
&=&\factor \pmatrix{-{10\over\sqrt{3}}&-{5\over 2}\sqrt{2\over 3}\cr
                       {5\over 2}\sqrt{2\over 3}&
                       \sqrt{10\over 3}\cr}(Mb)^2,\label{A7A8}\\
&&\\
{\bf A}_9
&=&\factor \pmatrix{0 & -5\sqrt{3\over 2} \cr
                       -5\sqrt{3\over 2} & 0 \cr}(Mb)^2,\label{A9}\\
&&\\
{\bf A}_{10}
&=&\factor \pmatrix{0 & 0\cr 0 & - 15\sqrt{7\over 15}}(Mb)^2,\label{A10}
\ee
where the matrices ${\bf A}_i$ (i=1,\ldots,10) act on the $j=1/2$ and $j=3/2$
spaces with the 11 (first) element of the matrices corresponding to the
diagonal matrix element for $j=1/2$.
The single nucleon isospin factor $<I_T> = {\sqrt{3}}$.
$b$ is the oscillator parameter for the shell model orbits.

\setcounter{equation}{0}
\setcounter{footnote}{0}
\renewcommand{\thefootnote}{\thesection.\arabic{footnote}}
\section{Semi-phenomenological R--matrix treatment for the A = 8
system}

\subsection{R--Matrix theory for $\beta$--decay processes}

Following Appendix D, the T--matrix for the $\beta$--decay processes is found
to
be
\be
T_{fi} &=& \sum {\Gamma_{2\alpha,n}(E_{2\alpha})
                  \bra{n;e\nu}H_w\ket{\phi_i}\over
                   E_{2\alpha} - E_n - R_n(E_{2\alpha})}.\label{Tmatrix41}
\ee
$R_n(E)$ is complex, namely
\be
   R_n(E) &=& D_n(E) - i I_n(E),\label{Rnexpression}
\ee
where the real functions $D_n(E)$ and $I_n(E)$ are the level shift  and
the half width\footnote{The width of the nth
level is defined as the value of $I_n(E)$ at the energy where
$E-E_n-D_n(E)=0$.} of the nth level, respectively.
$\Gamma_{2\alpha,n}(E)$ is the vertex function that connects
the nth level resonance to the two $\alpha$ states. The result given in
Eq.  \ref{Tmatrix41} is formally exact, provided $\ket{n}$ is a complete
set in the Hilbert space of interest. All effects of
$\Delta V$ are represented by the functions $R_n(E)$ and
$\Gamma_n(E)$. We shall not use a complete set of $\bra{n}$; we only need a
few of the most important shell model states, as shown in Appendix D. In
addition, only the (strong and electromagnetic) interaction effects at short
distances ($r<5fm$) are included in $R_n(E)$ and $\Gamma_n(E)$. The residual
effects are written as
\be
T_{fi}=\sum_{shell \, model}(\ldots) + \delta T_{fi},\label{Tfidel}
\ee
where residual part $\delta T_{fi}$ will be specified in the following.

 The imaginary part of $R_n(E)$ is related to the
magnitude of $\Gamma_n(E)$ through the optical theorem \cite{GWscat}.
In the particular energy range (0 $<$ E $<$ 16 MeV)
considered here, the only
open channel into which the shell model states $\ket{n}$ can decay
is the $2\alpha$ one. The optical theorem requires that

\be
   I_n(E) &=& \pi \rho(E) |\Gamma_{2\alpha,n}(E)|^2,
\label{OpTheorm1}\\
   \rho(E) &=& {1\over 2} M_{\alpha}\sqrt{M_{\alpha} E},\label{Densstate}
\ee
where we have assumed the center of mass (c.m.) frame for $^8$Be$^*$ and used a
non--relativistic approximation
\be
  E &=& {k^2\over M_\alpha},\label{Ek}
\ee
with $k$ the relative momentum between the two $\alpha$ particles.

At low energies, the R--matrix parameters $I_n(E)$ and
$\Gamma_{2\alpha,n}(E)$ are independent of the details of the interaction
potential $\Delta V$.  Their asymptotic behavior can be derived from Eqs.
\ref{Gammaalphan}, and \ref{LSeqforFn}. Namely, if $\Delta V$ has a short range
(the long range Coulomb interaction between two $\alpha$ particles is treated
separately in this study) and is less singular
than\footnote{Albeit
the definition of $r$ can be ambiguous if the $\alpha$ particle is
composite and the overlap between the two $\alpha$ particles is strong,
the singularity
of the effective $\Delta V$ can be smoother than the one extracted from
$\Delta V$ at larger distance where the overlap between these particles are
negligible due to the compositeness of the particles. Therefore the statement
based on point particle pairs given here is also expected to be
true for composite particles.} $1/r^{2-\epsilon}$ ($\epsilon<<1$)
as $r\to 0$, then
\be
 \ket{2\alpha}&\stackrel{|{\bf k}|\to 0} \sim &
                        (a|{\bf k}|)^J,\label{psilimit}\\
\ket{\psi_n} &\stackrel{|{\bf k}|\to 0} \sim &
                        constant,\label{psinlimit}
\ee
where $a$ is some scale factor of dimension length that is roughly
the size of $\Delta V$
and $J$ is the total angular momentum of the state with
two asymptotic $\alpha$ particles as $r\to\infty$.
By combining Eqs. \ref{psilimit},  \ref{psinlimit} and  \ref{Gammaalphan}, one
gets
\be
\Gamma_{2\alpha,n}(E) &\stackrel{E\to 0}\sim& E^{J\over 2},\label{Gammalimit}
\ee
where use has been made of the non--relativistic spectrum for
the $\alpha$ particles Eq.  \ref{Ek}.

  From Eq.  \ref{OpTheorm1}, we get
\be
 I_n(E)&\stackrel{E\to 0}\sim& E^{J+{1\over 2}}.\label{Inlimit}
\ee
At low energies, $D_n(E)$ and $I_n(E)$ are expected to be smooth
functions of $E$. They can be expanded in a power series of $E$ with the
expansion coefficients treated as parameters. These parameters are
determined from experimental data for $\alpha$--$\alpha$ scattering phase
shifts and allowed $\beta$--decay rates in the $A=8$ system.

We shall require that $D_n^J(E)$ is analytic in the energy region of interest.
The denominator $E-E_n^J-D_n^J(E)$ can therefore be written as
\be
E-E_n^J-D_n^J(E) &=&
E-E^{0,J}_n-a(E-E^{0,J})^2-b(E-E^{0,J})^3+\ldots,\label{Dexpansion}
\ee
with $E^{0,J}_n$ the experimental resonant energy of the nth level of
total angular momentum J. Terms of higher power in $E-E^{0,J}_n$ represented
by ``\ldots'' are expected to be unimportant around resonant energies.
The parameters $a$ and $b$ are to be determined in the following.

Next, let's consider the vertex function. From the low energy
behavior given by Eq.  \ref{Gammalimit},
$\Gamma_n^J(E)$ can be written as
\be
\Gamma_n^J(E) &=& N_0 E^{J\over 2} (1 - c E + \ldots),\label{GammaExpan}
\ee
with ``$\ldots$'' representing higher order terms in powers of $E$ that can
be ignored at low energies, and $N_0$, c are coefficients.
The half width function $I^J_n(E)$ is related to $\Gamma_n^J(E)$ through
the optical theorem, Eq.  \ref{OpTheorm1}. In the following, we shall
express the magnitude of $\Gamma_n^J(E)$, in terms of $I_n^J(E)$ and
parameterize $I_n^J(E)$ as
\be
I_n^J(E) &=& I_n^{0,J}\sqrt{E\over E_n^{0,J}}
         \left ({E\over E_n^{0,J}}\right )^J
         {(1-cE)^2\over (1-cE_n^{0,J})^2},\label{IJnExpan}
\ee
where $I_n^{0,J}$, and c are coefficients to be determined. The vertex
function $\Gamma_n^J(E)$ is assumed to be real function of $E$ on the
real $E$--axis.

\subsection{$\alpha$--$\alpha$ scattering}

    The R--matrix theory can also be applied to the scattering of
two $\alpha$ particles. The collision of two $\alpha$ particles can
be viewed as proceeding through various resonant levels of the $A=8$ system,
plus potential scattering between the two $\alpha$ particles at
larger distances where the overlap between the two $\alpha$ particles
is small.  The rigorous relationship between the R--matrix
parameters in the $\beta$--decay processes and the resonant scattering
cross section of the $\alpha$ particles is
indirect. Thus, a more explicit ansatz has to be made in order to relate
the $\beta$--decay and the $\alpha$--$\alpha$
scattering processes in a way that is convenient for the
phenomenological applications.

We assume that the scattering phase shifts of the $\alpha$ particles
due to interactions at short distances (the resonant scattering), $\delta
_J^{(res)}
(E)$, and at
large distances
(the potential scattering), $\delta _J^{(pot)}(E)$, are additive.
The resonant part of the T--matrix for the collision
of the two $\alpha$--particles in a state of total angular momentum $J$
can be
related to phase shifts generated by the resonant states and by
the potential scattering as
\be
t_J(E) &=& -{2\over\pi M_\alpha k}
e^{i\delta_J(E)}sin\delta_J(E),\label{alTmatr}\\
\delta_J(E) &=& \delta_J^{(res)}(E) + \delta_J^{(pot)}(E),\label{PhaseShifts}
\label{ResTmatr}
\ee
We shall assume further that the vertex function in the R--matrix does
not contain any
contribution from the potential scattering at large distances,
and that the potential scattering is provided solely by the
long range Coulomb interaction between the two $\alpha$ particles.
The phase shift due to the potential scattering is expressed as
\be
tan\delta_J^{(pot)}(E) &=& - F_J(rk)/G_J(rk),\label{deltasc}
\ee
where $F_J(x)$ and $G_J(x)$ are the regular and irregular Coulomb
functions \cite{FJGJ}
for the two $\alpha$--particle system,
and $r$ is the distance between the
two $\alpha$ particles where their relative radial wave functions
vanish when $\Delta V=0$.
It follows that there are infinite number of discrete values of $r$
that lead to
the same potential phase shift given by Eq.  \ref{deltasc}; all
of them are energy dependent. The energy dependence of r is expected
to be smaller when it is close to the origin, especially when the interaction
between the two $\alpha$ particles is stiff enough at small distances
that it can be treated as a hard core. The energy dependence of $r$ is thus
expected to be minimized by choosing it fairly close to the origin
\footnote{Since we can extrapolate the Coulomb
wave function given by a combination of $F_J$ and $G_J$ at large distances
all the way to  points close to the origin where the strong
interaction dominates,
the value of r is not necessarily
restricted to the region where the Coulomb interaction is most important.
In fact,
any value of r that gives the same energy dependence for the phase
shifts is acceptable. The meaning of r as a measure of the range
of the potential is meaningful only if the potential indeed contains a
stiff piece (e.g., hard core). In this case, a value of r that is located
near the hard core should be almost energy independent. Given the energy
dependence of $\delta^{(pot)}_J$, it must be emphasized
that, in the region where the strong interaction is important, the zeros
of the true wave function of the system are not located at the ones satisfying
Eq.  \ref{deltasc}}.

The T--matrix corresponding to the interactions that are responsible for
the resonant states (see Fig. 5)
in the $A=8$ system can be expressed in terms of
the R--matrix parameters, namely
\be
t_J^{(res)}(E) &=& -{2\over\pi M_\alpha k}
e^{i\delta^{(res)}_J(E)} sin\delta^{(res)}_J(E),\nonumber\\
&=& \sum {|g^J_{2\alpha,n}(E)|^2\over E-E^J_n-D^J_n(E) + iw^J_n(E)},
\label{alfascattering}
\ee
where it is assumed that the level shift function $D^J_n(E)$ is
the same as that in the R--matrix.
The relationship between $g^J_n(E)$
and $w^J_n(E)$ can be obtained by unitarity considerations of the
$\alpha$--$\alpha$ T--matrix; on the other hand,
the relationship between $w^J_n(E)$ and $I^J_n(E)$ is not obvious.
$\Gamma^J_n(E)$ is the
half width of the corresponding resonant level or the inverse
life time of that level. The decay of the level to two
$\alpha$ particles can
be pictured as that of a prepared state, namely as the eigenstate of the
shell model Hamiltonian $H_{shell}$, tunneling through the potential
``barrier'' $\Delta V$. From the time reversal invariance
consideration, it follows that in case of resonant
$\alpha$--$\alpha$ scattering, the time needed for the $\alpha$ particles
to tunnel into the potential ``barrier'' $\Delta V$ to form a well
prepared resonant state (corresponding to a shell model eigenstate) is
the same as the time it takes the same state to tunnel through the
potential ``barrier'' $\Delta V$ on its way out.
Thus the time delay due to the $\alpha$ particle staying in a resonant state
in $\alpha$--$\alpha$ scattering is twice as long as
the time for the corresponding shell model eigenstate to decay near a
resonant energy of that state. This qualitative argument is formally,
albeit not explicitly,
derived in Ref. \cite{GWscat}. From these considerations, it is assumed that
\be
w^J_n(E) = {1\over 2} I^J_n(E).\label{wIrelation}
\ee
As will be shown in the following, this relation leads to an excellent
synthesis between the $\beta$--decay $\alpha$ spectra and
$\alpha$--$\alpha$ scattering within the R--matrix formalism and without
the need of any ``intruder'' states.

Since a two spin zero particle state with definite total angular
momentum J has only one phase shift at a given energy,
the phase for each of the terms in the above sum has to be the same,
which allows us to write Eq.  \ref{alfascattering} as
\be
tan\delta_J^{(res)}(E) &=& -
\sum_{n} {w^J_n(E)\over E-E^J_n-D^J_n(E)}.\label{tandelres}
\ee
Together with Eq.  \ref{deltasc}, this equation yields
\be
\delta_J(E) = -tan^{-1}\sum_n {w^J_n(E)\over E-E^J_n-D^J_n(E)} - tan^{-1}
{F_J(rk)\over G_J(rk)}.\label{fullphaseshift}
\ee
Before going on to further specifications of $I^J_n(E)$ and $D^J_n(E)$,
it is useful to write the T--matrix for the $\alpha$--$\alpha$ scattering
in the total angular momentum J state as
\be
-{\pi\over 2} M_\alpha k t_J(E)
&=& e^{i\delta_J^{(res)}(E)+i\delta_J^{(pot)}(E)}\nonumber\\
&&\left [ sin\delta_J^{(res)}(E) cos\delta_J^{(pot)}(E)
        +cos\delta_J^{(res)}(E) sin\delta_J^{(pot)}(E)\right ].\nonumber\\
&&\label{TmatrxiSep}
\ee

The parameter ``r'' in the Coulomb functions $F_J(rk)$ and
$G_J(rk)$ is also expected to be energy dependent. The form of its energy
dependence is not easy to determine from known general properties of the
system.  So we shall take the following trial form
\be
   r(E) &=& r_0 (1 - z E^2),\label{RadiusExpans}
\ee
with ``z'' a parameter to be determined. It leads to a good fit
to the experimental data.

\noindent
\underline{$J^P = 0^+$ states:}

The $J^P=0^+$ states are shown in Fig. 3. The ground state of
$^8$Be$^*$ is
in the $J^P=0^+$ series and is rather narrow.
The next two states in this series are  broad and have
resonant energies at $E=20.2 MeV$ and $E=27.49 MeV$ respectively.
Since these states are well
separated from the ground state, it is justifiable to include only the
ground state in the R--matrix at low energies ($E^{cm}\le 12$ MeV).

The best fit (least square fitting) to the experimental $\alpha$--$\alpha$
phase shifts \cite{PhaseShift} in the $J^P=0^+$ state is achieved by using the
set of parameters shown in Table  \ref{J0para}. The result of the fitting
procedure
is shown in Fig. 6. The value of the $J^P=0^+$ phase shift at
$E=0.4 MeV$ given in Ref. \cite{PhaseShift} is interpreted as $180^o\pm 0.5$
instead of $0^o\pm 0.5$ in order to take into account the fact
that the $0^+$ resonance is at about $E=0.2 MeV$, which is smaller
than  $0.4MeV$.

\noindent
\underline{$J^P=2^+$ states:}

The $J^P=2^+$ states are our prime interest. Fig. 3 shows
a series of experimental low lying states with $J^P=2^+$. In the
energy region considered here, we shall assume that only the lowest
three states are important, namely the states with energy $E=3.04$ MeV,
$16.63$ MeV and $16.92$ MeV.
Higher states are assumed to have negligible influence for $\sim 12 MeV$.
The state at $E=3.04$ MeV
is an isosinglet state. The nearly degenerate doublet states at an energy
near 16 MeV are mixtures of isosinglet and isotriplet.  The latter component
does not
contribute to the
$\alpha$--$\alpha$ scattering to lowest order in the fine structure
constant $\alpha=e^2/\hbar c$. The
mixing angle between the isosinglet and the isotriplet
is nearly $45^0$ \cite{IsoMixing1,IsoMixing2}.
This mixing angle will be determined by fitting the
$\alpha$ particle spectra in the $\beta$--decay processes. They are not
relevant
to the phase shift analysis\footnote{
This conclusion follows because the $\alpha$--$\alpha$ phase shift
expressed by the R--matrix contains no information about the resonance
other than its width and the location. Without further assumptions,
the isospin mixing of the resonant states is not easily determined by
the phase shift analysis in the present approach.}. In addition,
the widths of these two states have
very small influence on the $\alpha$--$\alpha$ phase shift in the energy region
considered in Fig. 7.

The best least square fit to the experimental $\alpha$--$\alpha$
phase shifts \cite{PhaseShift}
in the $J^P=2^+$ state is achieved by using the set of parameters
shown in Table  \ref{J2para}, in which the
widths of the two $16$ MeV doublet states are not determined from the
$\alpha$--$\alpha$ scattering phase shift but rather from the
$\alpha$--spectra in the $\beta$--decay processes.
The result of the fitting is shown in
Fig. 7.

It can be noted that the value of r obtained in the $J^P=2^+$ channel
is smaller than two times the $\alpha$ particle size. This feature should
not be regarded as a shortcoming of the procedure.
As is pointed out in section 3.a, the value of $r$ is not necessarily
restricted to be outside of the region where the $\alpha$--$\alpha$ overlap is
unimportant. In our approach, $r$ represents the location of
a zero of the radial wave function before the decay potential $\Delta V $ is
turned on. From this point of view, there are many equivalent
r's.

\noindent
\underline{$J^P=4^+$ state:}

The $J^P=4^+$ series of states are also shown in Fig. 3.
The lowest state is at 11.4 MeV. Higher states in the $J^P=4^+$ series will
be neglected in the following.

The best least square fit to the experimental \cite{PhaseShift}
$\alpha$--$\alpha$ phase shifts
for $J^P=4^+$ is achieved by using the set of parameters shown
in Table  \ref{J4para}. The result of the fitting procedure
is shown in Fig. 8.

\subsection{The $\alpha$ particle spectra in the $\beta$--decay processes
         of the A=8 system}

The unpolarized differential $\beta$--decay rate follows from
Eqs. \ref{dWe} and  \ref{dW0}, namely
\be
{dW\over dE_r} &=& {2\over 5\pi^2} G_F^2 cos^2\theta_c |g_A|^2
M_\alpha \sqrt{M_\alpha E_r} |A_2(E_r)|^2 D(E_r),\label{UnpolRate}\\
D(E_r) &=& \int_{m_e}^{E_0-E_r} d\epsilon_e\epsilon_e\sqrt{\epsilon_e^2
-m_e^2} \nonumber \\ &&
\left (E_0-E_r-\epsilon_e\right )^2 F(Z,\epsilon_e) R^{(0)}_1
(E_r,\epsilon_e),\label{Dfunction}\\
F(Z,\epsilon_e) &=&{2\pi\eta\over e^{2\pi\eta}-1},\label{FZC}\\
\eta &=& \mp {Z\alpha\epsilon_e\over \sqrt{\epsilon_e^2-m_e^2}},\label{FandEta}
\ee
with $Z=4$ and $A_2(E_r)$ the energy dependent matrix element obtained
from the shell model amplitude Eq.  \ref{A2} using the R--matrix formalism
(see the following).

 From Eq.  \ref{Tmatrix41}, it follows that the matrix element between the
initial $^8$Li or $^8$B state and the final resonant state $^8$Be$^*$
that eventually decays to two $\alpha$ particles can be
related to a set of shell model matrix elements between the same initial state
and a discrete set of shell model $^8$Be$^*$ states through
\be
\braa{J_f}O_J\kett{J_i} &=& e^{i\delta_{J_f}(E)}
\sum{\Gamma_n^{J_f}(E_r)\braa{J_f;n}O_J\kett{J_i}
       \over
       \sqrt{[E_r-E_n^{J_f}-D_n^{J_f}(E_r)]^2 + I_n^{J_f}(E_r)^2}},
\label{RmatrixMat}
\ee
where $\braa{J_f}O_J\kett{J_i}$ denotes any reduced matrix elements, and
the index ``n'' enumerates different shell model states with the same parity
and spin (total angular momentum) $J_f$. The common phase factor for
each of the terms in the summation is written as a multiplicative  factor
of the whole expression on the r.h.s. of Eq.  \ref{RmatrixMat} with
the vertex function $\Gamma_{J_f}^n(E)$ real. This expression is, in
principle,
exact provided a complete set of shell model states is used and
the energy dependent vertex function $\Gamma_n^J(E)$, level
shift $D_n^J(E)$ and half width $I_n^J(E)$ are calculated from
Eqs. \ref{RnE},  \ref{Gammaalphan} and  \ref{LSeqforFn}. Due to the lack of
knowledge of the decay potential $\Delta V$ and the complexity of
the problem, a microscopic calculation of $\Gamma_n^J(E)$, $D_n^J(E)$
and $I_n^J(E)$ is beyond the scope of this work.  Instead, we shall
use the same parameterization of the R--matrix as in the
$\alpha$--$\alpha$ scattering case. Consistent with the treatment of the
$\alpha$--$\alpha$ scattering phase shift, only the 2$^+$ shell model
states at E=3.55 MeV, E=16.19 MeV and E=13.37 MeV
are used to evaluate the shell model transition amplitude. The
energies of these states are taken
from the fit of the $\alpha$--$\alpha$ phase shifts.  (This is reasonable,
since it is expected that there are level shifts for the shell model
states after switching on the
decay potential $\Delta V$). The shell model state lying
at E=17.45 MeV is not included in the sum of Eq.  \ref{RmatrixMat}.
Two additional parameters
in the $\beta$--decay processes that do not appear in the description of
$\alpha$--$\alpha$ scattering processes are needed.
The first parameter
is the isospin mixing angle between the isospin doublet members
at $E\sim 16$ MeV.
We shall define the mixing angle as
\be
\ket{E=16.63} &=& cos\phi \ket{T=1} + sin\phi \ket{T=0},\label{MixIso1}\\
\ket{E=16.92} &=&-sin\phi \ket{T=1} + cos\phi \ket{T=0},\label{MixIso2}
\ee
where $\ket{T=0}$ and $\ket{T=1}$ denote the isosinglet and isotriplet
shell model states respectively. Since only the T=0 components within each
state of the doublet can decay into two $\alpha$ particles, the contribution
of the T=1 part of  each state is not included in the R--matrix. In addition,
there are relative signs between $\Gamma^{J=2^+}(3.04)$,
$\Gamma^{J=2^+}(16.63)$
and
$\Gamma^{J=2^+}(16.92)$ that cannot be fixed from the sum rule relation
Eqs. \ref{OpTheorm1} and  \ref{Densstate}. They have to be determined in
the fitting. The second parameter is related to the inclusion of the potential
scattering in the R--matrix formalism for the $\beta$--decay processes
consistent with the similar development for the $\alpha$--$\alpha$
scattering. This parameter is necessary, since in the fitting of the
$\alpha$--$\alpha$
scattering phase shift, it is natural to separate the potential scattering
between the two $\alpha$ particles at long
distances
and the confining strong and Coulomb resonant interaction at short distances
(characterized by the parameter $r$). Similar to Eq.  \ref{TmatrxiSep} we shall
write the T--matrix in the presence of the potential scattering between
the $\alpha$ particles as
\be
T(E) &=& e^{i\delta_J^{(pot)}(E)}\left [
 cos\delta^{(pot)}_J(E)T_0(E) + sin\delta^{(pot)}_J(E) \tilde T_0(E)
\right ].
\label{BetaTmatrix}
\ee
 $T_0(E)$ has the form of Eq.  \ref{Tmatrix41} but with the full
vertex function $\Gamma_n^J(E)$, the level shift function $D_n^J(E)$,
and the half width function $I_n^J(E)$ replaced by the ones obtained
from the phase shift fitting procedure outlined in the previous section.
$\tilde T_0(E)$ is given by
\be
\tilde T_0(E) &=& C\sum {\Gamma_n^J(E)\over I_n^J(E)}{
     E-E^J_n-D_n^J(E) \over
      E-E^J_n-R_n^J(E)}\bra{n;e\nu}H_w\ket{\phi_i},\label{tT0E}
\ee
with constant $C$ a parameter to be determined. The magnitude of $C$
is a rough measure for the direct weak transition from the initial
$^8$Li or $^8$B to the two $\alpha$ scattering states.

The two new parameters $C$ and $\phi$, along with the widths of
the isodoublet states
are adjusted so that a fit to the $\alpha$
particle spectra for the $\beta^+/\beta^-$ decay in $A=8$ system
is achieved. The resulting values for $C$ and $\phi$ are
\be
  \phi &=& 51.8^o,\label{IsoMixAngle}\\
  C  & = & 5.7\times 10^{-3}.\label{ParaC}
\ee
The widths of the doublet state are
\be
  w_0(16.63) &=& 0.206 MeV,\label{W63}\\
  w_0(16.92) &=& 0.303 MeV.\label{W92}
\ee
In addition, relative to $\Gamma^{J=2^+}(3.04)$, the signs of
$\Gamma^{J=2^+}(16.63)$ and
$\Gamma^{J=2^+}(16.92)$ are negative. Before ending this subsection, let's
compare the isospin mixing angle obtained here with the one obtained in
Ref. \cite{IsoMixing2}, namely $\phi = 50.3^o$.
Thus, the value for the isospin mixing angle obtained here agrees well with
that of Ref. \cite{IsoMixing2}.

The results of the theoretical calculation, with the nucleon weak form factors
given in Table 7, are compared  to the
experimental \cite{Wilkinson}
data in Figs. 9 and 10.
The dashed  curve represents a calculation without the final state
potential scattering, namely $\delta^{(pot)}=0$ and the solid
curve is the full calculation. Figs. 11 and 12 are linear scale curves
corresponding to Figs. 9 and 10.

In obtaining the excellent fit to the $\alpha$ particle spectra
shown in Figs. 9 and 10, an upward energy shift of
0.17 MeV has been made to
 the theoretical expression Eq.  \ref{UnpolRate}.
It is as if only the lowest resonant energy
$J^P=2^+$ state were shifted upward by roughly 0.17 MeV.
This shift spoils the agreement between theory and the
experiment both in the shape of the $\alpha$ spectra and the phase shift.
Furthermore it is much too large to be attributable to an experimental energy
calibration \cite{Adelberger1}.  We do not understand its origin, only its
necessity.
In Warburton's fit to the $\alpha$--$\alpha$ scattering phase shift and the
$\alpha$ particle spectra of the $\beta$--decay processes of the A=8
system, a shift in the resonant energy of order 0.07 MeV was also
required \cite{Warburton}.
Henceforth we shall use the R--matrix development
without this shift.

It may be worth mentioning that the Barker and Warburton R--matrix
treatment, which is different from the one used here, obtains an equally
good (if not better)  fit to the
%$\alpha$--$\alpha$ scattering
%phase shift and the
$\alpha$--spectra \cite{Warburton} without
the need of low lying ``intruder'' states if the value of $R_c$ is
chosen to be
around 4.5 fm. Their R--matrix formalism does not, however, allow a
straightforward and consistent
extension to the forbidden transition matrix elements.
On the other hand, in the R--matrix
developed here, the allowed and forbidden transition amplitudes are treated
on an equal footing, so that the study of PTRIV phenomenae,
which are intrinsically due to forbidden processes, can be studied with
 the R--matrix determined in the previous sections. This kind of
 ``predictive power'' is the main reason that we adopted the present approach.

\setcounter{equation}{0}
\setcounter{footnote}{0}
\renewcommand{\thefootnote}{\thesection.\arabic{footnote}}
\section{PTRIV due to the final state interactions
between two final $\alpha$ particles}

\subsection{Energy dependence of the PTRIV response functions}

 Eqs. \ref{Rt3}--\ref{Rt6} show that the dependence of
$R_1^{PTRIV}$ and $R_2^{PTRIV}$ on the energy of the charged lepton
energy $\epsilon_e$ is quadratic. However, their dependence on the two
$\alpha$ particle relative energy is not trivial.
First there are strong $E_r$ dependence of $\eta_6(2^+,0^+)$,
$\eta_9(2^+,0^+)$, $\eta_6(2^+,4^+)$ and $\eta_9(2^+,4^+)$.
Figs. 13--16 give the energy dependence of
these quantities.
Functions $\eta_6(2^+,0^+)$ and $\eta_9(2^+,0^+)$ are dominated
by the $J^P=0^+$ resonance located at $E=0$; they have a zero near
$E\sim 3$ MeV. $\eta_6(2^+,4^+)$ and $\eta_9(2^+,4^+)$ are dominated
by the $J^P=4^+$ resonance located at $E\sim 11$ MeV; they do not
possess any zero in the energy range 2 -- 5 MeV. Also, they depend on
factors
$sin(\delta_2-\delta_0)$ and $sin(\delta_2-\delta_4)$, which
originate from the phase factors in the wave functions (see
Eqs. \ref{BeFinal} and  \ref{BetaTmatrix}). There
is an energy near $E\sim 3$ MeV where the $J^P=0^+$ phase shift, $\delta_0$,
is equal to the the $J^P=2^+$ state phase shift as can
be seen by a comparison of Fig. 6 and Fig. 7.
Therefore, there is a zero in $sin(\delta_2-\delta_0)$.
However, $sin(\delta_2-\delta_4)$, does not have a zero in this energy range.

In a test of  genuine time reversal invariance, which is associated with
the time reversal invariance of the underlying Lagrangian,
experiments should be designed to select events in which the PTRIV
contributions are small.
In Figs. 17 and 18 the value of $R_1^{PTRIV}$ and
$R_2^{PTRIV}$ are plotted against $E_r$ and $\epsilon_e$
for the $\beta^+$ decay of A=8 system. It can be seen that both
of $R_1^{PTRIV}$ and $R_2^{PTRIV}$
pass through zero for several combinations of $E_r$ and $\epsilon_e$,
along a line located at $E_r\sim 3$ MeV. $R_1^{PTRIV}$ is small at low
$e^\pm$ energies $(\epsilon_e)$ and increases with $\epsilon_e$; it changes
more
quickly with $E_r$. $R_2^{PTRIV}$ changes less quickly
at $\epsilon_e\approx 5.5$ MeV than at other values of $\epsilon_e$
and possesses  zeros somewhere near
$E_r\approx 3$ MeV and $\epsilon_e\le 8$ MeV.
Figs. 19 and 20 are
3--dimensional plots of the values of $R_1^{PTRIV}$ and $R_2^{PTRIV}$ against
$E_r$ and $\epsilon_e$ for the $\beta^-$ decay in the A=8 system.
The general features of $R_1^{PTRIV}$ are the same as those for the $\beta^+$
decay. $R_2^{PTRIV}$ is somewhat different from the corresponding
$\beta^+$ decay due to the $v-a$ interference effects.
There still are zeros around $E_r\approx 3$ MeV
at different $\epsilon_e$, but $R_1^{PTRIV}$ turns negative at high
values of $E_r$. It changes slower with $E_r$ when $\epsilon_e\approx 5.5$
MeV compared with its value at other values of $\epsilon_e$
in the energy region (of $E_r$) considered.

\subsection{PTRIV in case the charged lepton energy is not measured}

In case the charged lepton energy is not measured, an average over its
energy distribution has to be made. From Eq.  \ref{Dfunction}, it is
natural to define the averaged PTRIV response function as
\be
\bar R_i^{PTRIV}(E_r) &=& {1\over N}\int^{E_0-E_r}_{m_e}
      d\epsilon_e\epsilon_e\sqrt{\epsilon_e^2-m_e^2}(E_0-E_r-\epsilon_e)^2
      F(E_r,\epsilon_e) R_i^{PTRIV}(E_r,\epsilon_e),\nonumber\\ &&
\label{avRPTRIV}\\
N &=& \int^{E_0-E_r}_{m_e}
      d\epsilon_e\epsilon_e\sqrt{\epsilon_e^2-m_e^2}(E_0-E_r-\epsilon_e)^2
      F(E_r,\epsilon_e),\label{NormalFac}
\ee
with $i=1,2$.

The $E_r$ dependence of the average PTRIV response functions are
shown in Figs. 21 and 22.

\setcounter{equation}{0}
\setcounter{footnote}{0}
\renewcommand{\thefootnote}{\thesection.\arabic{footnote}}
\section{$e^\pm$--nucleus Coulomb interaction and related PTRIV}

The treatment of the Coulomb effects in the $\beta$--decay processes
has a long history \cite{AllowedCoul1,AllowedCoul2,AllowedCoul3}.
However the coordinate space version of
the treatments are not easily adapted to the formalism developed in
this study. We shall use a somewhat different approach.

Although it is not necessary to consider the full
effects of the charged lepton--nucleus scattering for the purpose of this
paper, we nevertheless treat them here. The particular aspect of the
Coulomb scattering effects of the charged lepton studied in detail here is
the Coulomb scattering induced PTRIV effects in
the $\beta$--decay processes of the A=8 system. Similar effects were studied
by Jackson, Treiman and Wyld for general allowed $\beta$--decay
processes \cite{JTW2}. The Coulomb PTRIV effects were later related to
$e^\pm$--Coulomb phase shifts \cite{Brodine},
similar to the relationships developed in the previous
section. When we go beyond the non--relativistic limit, infinite
series of partial waves are required (see later sections of this chapter).
For our specific problem, we shall not start with
the spherical basis to calculate the contribution of the $e^\pm$--nucleus
Coulomb scattering to the PTRIV, but rather make use
of the integral form of the Dirac equation for the charged leptons. These
sets of integral equations are relativistic Lippmann--Schwinger equations,
with the boundary conditions built--in from the beginning.

\subsection{Modified weak interaction Hamiltonian due to the $e^\pm$--nucleus
         Coulomb scattering}

When the Coulomb scattering of the $e^\pm$ due to the presence of
the charged nucleus is considered, $\bf q$ can have a distribution of
values associated with different ${\bf  k}'_e$ at the point of the $e^\pm$
emission. If we take into account  the Coulomb scattering, the
weak Hamiltonian can be written as

\be
   H_W &=& {G_F\over\sqrt{2}}\int {d^3q\over (2\pi)^3}
         \bra{{\bf k}_e,{\bf k}_\nu}j_\nu({\bf q})\ket{0}
            J^\mu(-{\bf q}),\label{WHTcoulomb}
\ee
where the helicity indices for the leptons have been suppressed.

The matrix elements of the leptonic weak current operator in the
$\beta$--decay processes can be written as
\be
\bra{{\bf k}_e;{\bf k}_\nu}j_\mu({\bf q})\ket{0} &=&\int d^3x
e^{-i{\bf q}\cdot{\bf x}}\bra{{\bf k}_e;{\bf k}_\nu}j_\mu({\bf x})\ket{0}
\nonumber\\
&=& (2\pi)^3\delta({\bf k}'_e+{\bf k}_\nu+{\bf q})\nonumber\\
&&
\left\{
\begin{array}{cc}
\bar\psi^{(-)}_{e^-,{\bf k}_e}({\bf k}'_e)\gamma_\mu
(1-\gamma^5) v_{\bar\nu}({\bf k}_\nu)
& \mbox{for $\beta^-$ decay}\cr
\bar u_{\nu}({\bf k}_\nu)\gamma_\mu
(1-\gamma^5)\psi^{(+)}_{e^+,{\bf k}_e}({\bf k}'_e)
& \mbox{for $\beta^+$ decay}\cr
\end{array}\right.,
\label{LPCFourier}
\ee
with $u_{\nu}({\bf k})$ and $v_{\bar\nu}({\bf k})$ massless neutrino and
antineutrino spinors respectively. Eqs. \ref{WHTcoulomb}
and  \ref{LPCFourier} can be compared to earlier work on the
Coulomb corrections to the allowed $\beta$--decay in the
{\em elementary particle approach} \cite{CoulombW,HolsteinC,Sato}.
The Fourier transformation of the e$^\pm$ wave
functions are given by
\be
\psi_{e^-,{\bf k}_e}({\bf k}'_e) &=& \int d^3x e^{-i{\bf
     k}'_e\cdot{\bf x}} \psi_{e^-,{\bf k}_e}({\bf x}),\label{PsiemFour}\\
\psi_{e^+,{\bf k}_e}({\bf k}'_e) &=& \int d^3x e^{i{\bf
     k}'_e\cdot{\bf x}} \psi_{e^+,{\bf k}_e}({\bf x}).\label{PsiepFour}
\ee

\subsection{$e^\pm$ wave functions and their relativistic
Lippmann--Schwinger equations}

The wave functions for the $e^\pm$ obey Dirac equations for
relativistic charged fermions in Coulomb fields
\be
  (h_0 + V + \beta m_e)\psi_{e^-} &=& \eps_e\psi_{e^-},\label{ElectronEq}\\
  (h_0 - V + \beta m_e)\psi_{e^+} &=& \eps_e\psi_{e^+},\label{PositronEq}
\ee
where $V=-Z\alpha/|{\bf x}|$ in coordinate space and
\be
h_0 &=& -i\alpha\cdot\nabla + \beta m_e,\label{h0fore}
\ee
with $\alpha^i = \gamma^0\gamma^i$ and $\beta = \gamma^0$.  $h_0$ is the
free Dirac particle Hamiltonian. A very detailed study of the scattering
state solutions to Eq. \ref{ElectronEq} or  \ref{PositronEq} can be found
in the book by Greiner, M{\"u}ller, and Rafelski \cite{GMR}.

For our purpose, it is more convenient to use a relativistic
Lippmann--Schwinger equation for the $e^\pm$ wave functions, namely
\be
\psi^{(\pm)}_{e^-,{\bf k}_e}({\bf x}) &=& u({\bf k}_e) e^{i{\bf
      k}_e\cdot{\bf x}} + \int d^3x'
       \bra{\bf x}{1\over {\epsilon_e - h_0 \pm i\epsilon}}
        T_{e^-}^{(\pm)}\ket{{\bf x}'}u({\bf k}_e)e^{i{\bf k}_e\cdot{\bf x}'},
                  \nonumber\\ &&\label{Psieminus}\\
\psi^{(\pm)}_{e^+,{\bf k}_e}({\bf x}) &=& v({\bf k}_e) e^{-i{\bf
        k}_e\cdot{\bf x}} + \int d^3x'
        \bra{\bf x}{1\over {-\epsilon_e - h_0 \pm i\epsilon}}
        T_{e^+}^{(\pm)}\ket{{\bf x}'}v({\bf k}_e)e^{-i{\bf
          k}_e\cdot{\bf x}'}
          ,\nonumber\\ &&\label{Psieplus}
\ee
where $\epsilon_e\ge 0$.
The free Dirac Hamiltonian $h_0$ and the
matrix elements $\bra{\bf x}$ \ldots $\ket{{\bf x}'}$ are
$4\times 4$ matrices in Dirac space. In fact, Eq. \ref{Psieplus} corresponds
to the negative energy solution of Eq. \ref{Psieminus}, but in order to
use field theoretical language, $\epsilon_e$ is restricted to be
positive and equations for $e^-$ and $e^+$ are written separately.
Similar to the free Dirac spinor \cite{BDbook}, the positron wave function
can be obtained from the electron wave function by a charge
conjugation and a change of sign of the Coulomb potential, namely
\be
\psi^{(\pm)}_{e^+,{\bf k}_e} [V_{e^+}] &=&
C\gamma^0\psi^{(\mp)*}_{e^-,{\bf k}_e}[V_{e^+}],\label{WaveFuncForPos}
\ee
where $V_{e^+}$ is the Coulomb potential experienced by a positron and
$[V]$ denotes the potential used in solving the corresponding wave function.
Due to the long range nature of the Coulomb potential, the Lippmann--Schwinger
equation may not be well defined. This difficulty is overcome
by introducing a screened Coulomb field, which is the case in any realistic
experimental situation. The ``pure'' Coulomb case is realized in the limit of
an infinite screening length. In most of the following formal
discussions, we shall assume that the above limiting procedure is
taken.

The Dirac spinors $u({\bf k}_e)$ and $v({\bf k}_e)$
represent a free electron and positron respectively;
they satisfy
\be
(\sla{k} - m_e)u({\bf k}) &=& 0,\label{Uequation}\\
(\sla{k} + m_e)v({\bf k}) &=& 0.\label{Vequation}
\ee
$T_{e^-/e^+}^{(\pm)}$ is the T--matrix
of the $e^\pm$ in the Coulomb field of the nucleus under
study. $T_{e^-}^{(\pm)}$ and $T_{e^+}^{(\pm)}$ satisfy equations
\be
T_{e^-}^{(\pm)} &=& V + V{1\over{\epsilon_e-h_0\pm i\epsilon}}
T_{e^-}^{(\pm)},\label{Teminus}\\
T_{e^+}^{(\pm)} &=& (-V)+ (-V){1\over{-\epsilon_e-h_0\pm i\epsilon}}
T_{e^+}^{(\pm)}.\label{Teplus}
\ee
In momentum space, the
Lippmann--Schwinger equations for an electron and a positron in a
Coulomb potential become
\be
\psi_{e^-,{\bf k}_e}^{(\pm)}({\bf k}'_e) &=& \left [
(2\pi)^3\delta({\bf k}_e-{\bf k}'_e) +
{1\over \eps_e-\alpha\cdot{\bf k}'_e-\beta m_e\pm i\eps}
T_{e^-}^{(\pm)}({\bf k}'_e,
{\bf k}_e)\right ] u({\bf k}_e),\nonumber\\ &&\label{LSEQspim}\\
\psi_{e^+,{\bf k}_e}^{(\pm)}({\bf k}'_e) &=& \left [
(2\pi)^3\delta({\bf k}_e-{\bf k}'_e) +
{1\over -\eps_e+\alpha\cdot{\bf k}'_e-\beta m_e\pm i\eps}
T_{e^+}^{(\pm)}({\bf k}'_e,
{\bf k}_e)\right ] v({\bf k}_e).\nonumber\\ &&\label{LSEQspip}
\ee
Define the pair of operators
\be
O_{e^-}^{(\pm)} ({\bf k}'_e,{\bf k}_e) &=&
  {1\over
\eps_e-\alpha\cdot{\bf k}'_e-\beta m_e\pm i\eps}T_{e^-}^{(\pm)}({\bf k}'_e,
{\bf k}_e),\label{Ominus}\\
O_{e^+}^{(\pm)} ({\bf k}'_e,{\bf k}_e) &=&
  {1\over
 -\eps_e+\alpha\cdot {\bf k}'_e-\beta m_e\pm i\eps}
  T_{e^+}^{(\pm)}({\bf k}'_e,{\bf k}_e).\label{Oplus}
\ee
The conjugate operators are
\be
\bar O_{e^-}^{(\pm)}({\bf k}_e,{\bf k}'_e) &=&
 T_{e^-}^{(\mp)}({\bf k}_e,{\bf k}'_e){1\over \eps_e-\alpha\cdot{\bf k}'_e
-\beta m_e \mp i\eps},\label{Obarminus}\\
\bar O_{e^+}^{(\pm)}({\bf k}_e,{\bf k}'_e) &=&
 T_{e^+}^{(\mp)}({\bf k}_e,{\bf k}'_e){1\over -\eps_e+\alpha\cdot{\bf k}'_e
-\beta m_e \mp i\eps}.\label{Obarplus}
\ee
These operators will be used when the matrix elements of the leptonic
weak current operators are studied.

\subsection{The leptonic tensor}

Using operators given by Eqs. \ref{Ominus}--\ref{Obarplus}
the  leptonic tensor corresponding to the trace of the
bilinear products of the leptonic weak current operators becomes
\be
Trj^\mu({\bf q})j^{\nu\dagger}({\bf q}') &=& (2\pi)^6 \delta({\bf
q}+{\bf k}_e+{\bf k}_\nu)\delta({\bf q}'+{\bf k}_e+{\bf k}_\nu)L^{\mu\nu}
\nonumber\\
&&+(2\pi)^3\delta({\bf q}+{\bf k}_e+{\bf k}_\nu)A^{\mu\nu}({\bf q}')+
   (2\pi)^3\delta({\bf q}'+{\bf k}_e+{\bf k}_\nu)B^{\mu\nu}({\bf q})
\nonumber\\
&&+C^{\mu\nu}({\bf q},{\bf q}'),\label{Tracejmujnu}
\ee
where $L^{\mu\nu}$ is given by Eq. \ref{Lmn}. $A^{\mu\nu}$, $B^{\mu\nu}$ and
$C^{\mu\nu}$ are

\noindent
\underline{For the $\beta^-$ decay}
\be
A^{\mu\nu}({\bf q}') &=& 2 Tr O_{e^-}(-{\bf k}_\nu-{\bf q}',{\bf k}_e)
(\sla{k}_e+m_e)\gamma^\mu \sla{k}_\nu \gamma^\nu(1-\gamma^5),
\label{Amunu1}\\
B^{\mu\nu}({\bf q})&=& 2 Tr(\sla k_e + m_e)
\bar O_{e^-}({\bf k}_e,-{\bf k}_\nu-{\bf q})
\gamma^\mu\sla k_\nu\gamma^\nu(1-\gamma^5),
\label{Bmunu1}\\
C^{\mu\nu}({\bf q},{\bf q}') &=& 2 TrO_{e^-}(-{\bf k}_\nu-{\bf q},{\bf k}_e)
(\sla k_e + m_e)\bar O_{e^-}({\bf k}_e,-{\bf
k}_\nu - {\bf q})\gamma^\mu\sla k_\nu\gamma^\nu(1-\gamma^5).\nonumber\\ &&
\label{Cmunu1}
\ee

\noindent
\underline{For the $\beta^+$ decay}
\be
A^{\mu\nu}({\bf q}') &=& 2 Tr(\sla k_e-m_e)
\bar O_{e^+}({\bf k}_e,-{\bf k}_\nu-{\bf q}')
\gamma^\nu\sla k_{\nu}\gamma^\mu (1-\gamma^5),\label{Amunu2}\\
B^{\mu\nu}({\bf q}) &=& 2TrO_{e^+}(-{\bf k}_\nu -{\bf q},{\bf k}_e)
(\sla k_e-m_e)\gamma^\nu\sla k_\nu\gamma^\mu(1-\gamma^5),
\label{Bmunu2}\\
C^{\mu\nu}({\bf q},{\bf q}') &=& 2 Tr O_{e^+}(-{\bf k}_\nu-{\bf
q},{\bf k}_e)(\sla k_e - m_e)\bar O_{e^+}({\bf k}_e,-{\bf k}_\nu-{\bf q}')
\gamma^\nu\sla k_\nu\gamma^\mu (1-\gamma^5).\nonumber \\ &&
\label{Cmunu2}
\ee

\subsection{On--Shell Coulomb T--matrix to all orders in the fine
structure constant $\alpha$}

Rigorous expressions for $A^{\mu\nu}$,\ldots,$C^{\mu\nu}$ involving the
T--matrices (on--shell and off--shell) of the charged lepton in the hadronic
Coulomb field are  expected to be very complicated. Due to the smallness of the
fine structure constant $\alpha=1/137$, it is reasonable to adopt an
approximation in which only terms to the first order in $Z\alpha$
are considered. This however can not be done straightforwardly by truncating
the Lippmann--Schwinger equation, since the truncation
leads to infrared divergences for many of the PTRIV observables.
Even when a screened Coulomb
potential is used, the value of the T--matrix in the forward direction
cannot be reliably obtained using the lowest order one, which is real,
when the screening length is much larger than the hadronic scale.
In fact, an all order (in $\alpha$) calculation yields a rapidly
varying phase for the T--matrix in the forward direction that cancels
the infrared divergences in the physical observables. This phase can
be calculated analytically in the non--relativistic limit \cite{QM1}
(i.e., when the momentum of the charged lepton is much less than
its mass). Away from the non--relativistic limit, the
Coulomb T--matrix does not have a closed form. It is however possible
to express it in terms of the Coulomb phase shifts obtained in
Refs. \cite{Mott,Bethe} and more recently in e.g.
Ref. \cite{GMR}. The first order approximation
in $Z\alpha$ is made only at the final step of the calculation.
For the special case of the $\beta$--decay processes of the A=8 system,
the electron (positron) energy is much large than its rest mass. This
allows us to use the T--matrix in its extreme relativistic limit, where
helicity is conserved. In this limit, it is easy to show that the Dirac
structure of the T--matrix is very simple. It can be written as
\be
t &=& t_2\otimes 1,\label{TDirac}
\ee
with ``$1$'' an unit $4\times 4$ matrix.  The matrix elements of t
between states with definite helicities have the form
\be
t^{\sl ++}({\bf k}',{\bf k}) &=& cos{\theta\over 2} t_2(\eps,\theta,\phi),
\label{Tpphel}\\
t^{\sl --}({\bf k}',{\bf k}) &=& t^{\sl ++}({\bf k}',{\bf k}),\label{Tmmhel}\\
t^{\sl +-}({\bf k}',{\bf k}) & \sim & t^{\sl -+}({\bf k}',{\bf k})
=0,\label{Tpmhel}
\ee
where ``+'' represents helicity  ``up'' and ``--'' represents helicity ``down''
and ($\theta,\phi$) are the polar angles between ${\bf k}$ and ${\bf k}'$.
Quantity $t_2(\eps,\theta,\phi)$, which is independent of $\phi$,
is related to the Coulomb phase shifts
through \cite{Jacob}
\be
t_2(\eps_e,\theta,\phi) &=& {1\over\epsilon_e |{\bf k}_e|}
\sum_{\kappa=1}^{\infty} \left [e^{\pm i\delta_\kappa}sin\delta_\kappa-
        e^{\pm i\delta_{-\kappa+1}}sin\delta_{-\kappa+1})\right ]
        P'_{\kappa-1}(cos\theta),\nonumber\\ &&\label{T2expansion}
\ee
with $\delta_{\pm \kappa}$  Coulomb phase shifts of the doublet states with
total angular momentum $j=|\kappa|-{1\over 2}$ and with opposite
parities, $P'_L(x) = dP_L/dx$ and $P_L(x)$ is the Legendre polynomial.
Due to the helicity conservation of the T--matrix, the phase shifts of
states with the same total angular momentum and with opposite parities
are the same. Using the fact that
$\delta_\kappa=\delta_{-\kappa}$, Eq. \ref{T2expansion} can be written as
\be
t_2(\eps_e,\theta) &=& {1\over\epsilon_e |{\bf k}_e|}
\sum_{\kappa=1}^{\infty} \left [e^{\pm i\delta_\kappa}sin\delta_\kappa-
        e^{\pm i\delta_{\kappa-1}}sin\delta_{\kappa-1})\right ]
        P'_{\kappa-1}(cos\theta).\nonumber\\ &&\label{T2expansion2}
\ee
PTRIV
correlation observables depend on the moments of the real part of
the on shell Coulomb T--matrix if a first order approximation in
$Z\alpha$ is taken in the final results. These moments are defined as
\be
a_n &=& {|{\bf k}_e|^3\over Z\alpha\eps_e}\int_{-1}^1 dx x^n
Re[t_2(\eps_e,x)], \label{Tmoments}
\ee
with $x=cos\theta$. Taking a first order approximation in $Z\alpha$
for the Coulomb phase shifts $\delta_\kappa$, $a_n$
can be expressed as
\be
a_n &=& \sum_{\kappa=1}^{\infty} (1+(-1)^{n+\kappa})\Delta f_\kappa
        - n \sum_{\kappa=1}^{n} \int^{1}_{-1} dx x^{n-1} P_{\kappa-1}(x)
        \Delta f_\kappa,\nonumber\\ &&\label{firstorderan}\\
\Delta f_{\kappa\ne 1} &=& {1\over2\kappa(\kappa-1)} + {1\over 2(\kappa-1)!}
\left [ (\kappa-1)^2(\kappa-2)-\kappa (\kappa-1) + 2\epsilon (2-\kappa)
\right ],\nonumber\\ &&\label{Deltaf}\\
\Delta f_1&=&-{1\over 2}+\eps,\label{Dettaf1}
\ee
where $\epsilon=0.57721566\ldots$ is the Euler--Mascheroni constant and
the Coulomb phase shift obtained in Ref. \cite{GMR} has been used.
The first few useful moments are
\be
a_0 &=& c^{(+)},\label{aaa0}\\
a_1 &=& c^{(-)}+\left ({1\over 2}-\eps\right),\label{aaa1}\\
a_2 &=& c^{(+)}+1,\label{aaa2}\\
a_3 &=& c^{(-)}+\left ({4\over 3}-{8\over 5}\eps \right ),\label{aaa3}
\ee
with
\be
c^{(+)} &=& 2\sum_{\kappa=2,4,\ldots}^{\infty}\Delta f_\kappa,\label{cplus}\\
c^{(-)} &=& 2\sum_{\kappa=1,3,\ldots}^{\infty}\Delta f_\kappa.\label{cminus}
\ee
The numerical value of $c^{(+)}$ and $c^{(-)}$ are given by
$c^{(+)} = -9.942\times 10^{-2}$ and $c^{(-)} = 9.942\times 10^{-2}$.
It can be easily seen that all these moments diverge if a first order
(in $Z\alpha$) approximation for the Coulomb T--matrix of the $e^\pm$
is used\footnote{Some combinations of the moments remains finite even
in the first order approximation for the Coulomb T--matrix.}
due to the singularity of the first order
T at $x=1$. The results of the Coulomb
T--matrix for the $e^\pm$ to all order in $Z\alpha$
suppresses the divergence at $x=1$ and
results in finite moments $a_n$. In addition, it can be shown that,
except for some special combination of the moments, $a_n$ contains
contributions from states with all possible total angular momenta when the
mass of the $e^\pm$ is not much larger than its 3--momentum (i.e.
away from the non--relativistic limit). However,
in the extreme relativistic limit, the average effects of these
states, represented by $c^{(+)}$ for even n and $c^{(-)}$ for odd n, are small
and the sum of $c^{(+)}$ and $c^{(-)}$ seems to be vanishingly small
in a numerical evaluation of the corresponding series.

\subsection{Dispersive part of the Coulomb corrected leptonic weak current}

Only the dispersive parts of Eqs. \ref{Amunu1}--\ref{Cmunu2} will
be studied in this paper, since they alone contribute to the PTRIV signals.
In addition, to the first order in $Z\alpha$,
the dispersive parts of Eqs. \ref{Amunu1}--\ref{Cmunu2}
 originate from the imaginary part of the
free Dirac particle propagators
\be
{1\over \eps_e-h_0\pm i\eps}&=& {P\over \eps_e-h_0}\mp i\pi
\delta(\eps_e-h_0),\label{EleProp}\\
{1\over -\eps_e-h_0\pm i\eps} &=& {P\over -\eps_e-h_0}\mp i
\pi\delta(-\eps_e-h_0),\label{PosProp}
\ee
where ``P'' denotes principal value.

\noindent
\underline{For $\beta^-$ decay}
\be
 D_{\pm}^{\mu\nu}|_{disp}({\bf q},{\bf q}') &=&
{i\over\pi}|{\bf k}_e|Re [t^{e^-}_2({\bf k}_e,{\bf k}'_e)]
Tr(\sla k_e\gamma^0\sla k'_e \mp \sla k'_e\gamma^0\sla
k_e)\gamma^\mu\sla k_\nu\gamma^\nu (1-\gamma^5),\nonumber\\
&&\label{tildeAmn1}
\ee
\noindent
and

\noindent
\underline{for $\beta^+$ decay}
\be
 D_{\pm}^{\mu\nu}|_{disp}({\bf q},{\bf q}') &=&
-{i\over\pi}|{\bf k}_e|Re [t^{e^+}_2({\bf k}_e,{\bf k}'_e)]
Tr(\sla k_e\gamma^0\sla k'_e \mp \sla k'_e\gamma^0\sla k_e)\gamma^\nu
\sla k_\nu\gamma^\mu (1-\gamma^5),\nonumber\\
&&\label{tildeAmn2}
\ee
where $D_{\pm}^{\mu\nu}$ is related to $A^{\mu\nu}$ and $B^{\mu\nu}$
given in Eqs. \ref{Amunu1} and  \ref{Bmunu1} or Eqs. \ref{Amunu2} and
\ref{Bmunu2} through
\be
 D_{\pm}^{\mu\nu}({\bf q},{\bf q}') &=&{1\over (2\pi)^3}
\int { dk_e k_e^2 }
 \left [ A^{\mu\nu}({\bf q}) \pm B^{\mu\nu}({\bf q}')\right ]
\delta(\eps_e-\eps_k),\label{Atiddef},
\ee
with ${\bf q}= -{\bf k}'_e-{\bf k}_\nu$, ${\bf q}'=-{\bf k}"_e-{\bf k}_\nu$,
${ k}_e=|{\bf k}'_e|=|{\bf k}"_e|$ and $\eps_k=\sqrt{k^2_e+m^2_e}$.

   The symmetric (with respect to $A^{\mu\nu}$ and $B^{\mu\nu}$) and dispersive
parts of the final $e^\pm$--nucleus Coulomb
scattering contribution to the leptonic tensor given by
$D_{+}^{\mu\nu}({\bf q},{\bf q}')$ has opposite time reversal
transformation properties to the non--interacting part of $L^{\mu\nu}$
given in Table 4.  Therefore the T--odd kinematic functions
of rank 2 are given in Table  \ref{KineFuncCoul},
which can be compared to the ones given in Table  \ref{KineFunc}. The
antisymmetric (with respect to $A^{\mu\nu}$ and $B^{\mu\nu}$)
and dispersive parts of the  final $e^\pm$--nucleus Coulomb scattering
contribution to the leptonic tensor given
by $D_-^{\mu\nu}({\bf q},{\bf q}')$ has the same time reversal transformation
properties as the non--interacting part of $L^{\mu\nu}$.
Both of these
kinematic functions contain an extra variable $\hat{\bf k}'_e$ which
has to be integrated out.

\subsection{PTRIV correlation observables}

The dispersive contribution of the $e^\pm$--nucleus Coulomb scattering
amplitude
to the T--odd differential $\beta$--decay rate $dW$ of the A=8 system can
therefore be written as
\be
dW &=& 2\pi\delta(E_i-E_f-\eps_e-\eps_\nu){1\over 2}G_F^2 cos^2\theta_c
\nonumber \\ &&
 \int {d^3q\over (2\pi)^3}{d^3q'\over (2\pi)^3}
Tr\left [ J^{\mu\dagger}(-{\bf q}) J^{\nu}(-{\bf q}') \right ]
Tr\left [ j_{\mu}^{\dagger}({\bf q}) j_\nu({\bf q}')\right ]_{disp}.
\nonumber\\ &&\label{dWCoul}
\ee
By using the hadronic trace formula \cite{thesis}, and after some algebra
and integration over $d\Omega_{{\bf k}'_e}$,
it is found that in addition to Eqs. \ref{Rt1}--\ref{Rt6}, there are other
terms
that contribute to $dW^{PTRIV}$. Both $D_+^{\mu\nu}$ and
$D_-^{\mu\nu}$ contributes to the PTRIV correlation observables.
The PTRIV observable in the $\beta$--decay processes of the A=8 system
can be written as
\be
dW^{PTRIV} &=& dW_0 T_2(\hat {\bf k}):\nonumber\\ &&\left [\left (R_1^{PTRIV}
+R_{em,1}^{PTRIV}\right )\hat {\bf k}_e
\hat {\bf k}_e\times\hat {\bf k}_\nu + \left (
R_2^{PTRIV} + R_{em,2}^{PTRIV}\right )\hat {\bf k}_\nu
\hat {\bf k}_e\times\hat {\bf k}_\nu\right ],\nonumber\\
&&\label{dWPTRIV2}
\ee
where the tensorial contraction between the second rank Cartesian
tensor $T_2$ and two vectors by symbol ``:'' is defined by Eqs. \ref{T2AB}
and  \ref{T2ij}.  It is found that to order $\kappa^2$ in the hadronic
response function are
\be
R_{em,1}^{PTRIV} &=& \mp Z\alpha\left [
       {3\eps_e\over 8 M}\beta_1 \left(\pm f_c + f_M + f_T + {1\over 15}
       \sqrt{15\over 7}{\Delta\over M}f_1\eta_6\right )+\right.\nonumber\\
      && \left ({\eps_e^2\over
       M^2} \beta_2-{\eps_e(\Delta-\eps_e)
       \over  M^2} \beta_3\right )\times\nonumber\\
      && \left.\left (-{1\over 30}\sqrt{15\over 7}f_1\eta_6\mp{\sqrt
      2\over 20}\eta_8\mp{1\over 60}\sqrt{10\over 7}\eta_9 \mp
      {2\over 15}{1\over
      \sqrt 7}\eta_{10}\right )\right ],\label{Rem1}\\
R_{em,2}^{PTRIV} &=& \mp Z\alpha \left[{\eps_e (\Delta-\eps_e)\over 2M^2}
\beta_1
\left (\pm {12\over 5}\sqrt{5\over 21}f_1\eta_6 - {\sqrt{2}\over
15}\eta_8 + {1\over 5}\sqrt{10\over 7}\eta_9 - {4\over 35}
{1\over\sqrt{7}}\eta_{10}\right )\right.\nonumber \\ && +
 \left ({\eps_e(\Delta-\eps_e)\over M^2} \beta_3-
              {(\Delta-\eps_e)^2\over M^2} \beta_4\right )\times\nonumber\\
&&\left.\left (-{1\over 30}\sqrt{15\over 7}f_1\eta_6\mp{\sqrt 2\over 20}\eta_8
         \mp {1\over 60}\sqrt{10\over 7}\eta_9 \mp {2\over
         15}{1\over\sqrt 7}\eta_{10}\right )\right ],
\label{RPTRIV3}
\ee
where the reduced response functions $\eta_i$ are given in section 2, and
\be
\beta_1 &=& a_2-a_0 = 1,\label{bbb1}\\
\beta_2 &=& 3a_3+3a_2-a_1-a_0=6-{14\over 5}\eps,\label{bbb2}\\
\beta_3 &=& 3a_2+2a_1-a_0=4-2\eps,\label{bbb3}\\
\beta_4 &=& 2a_1+2a_0=1-2\eps.\label{bbb4}
\ee

In order to test whether or not the underlying Lagrangian of the
system is invariant under time reversal,
one should concentrate on events in which the PTRIV
contribution is small,  so that
the genuine TRIV can be extracted.
Figs. 23 and 25 are
3--dimensional plots of $R^{PTRIV}_{em,1}$ and $R^{PTRIV}_{em,2}$ for
the $\beta^+$ decay process of the A=8 system. $R^{PTRIV}_{em,1}$ for
the $\beta^+$ decay process, which
contains the first forbidden contributions (namely contributions of order
$\kappa/M$) obtained by
Holstein \cite{HolsteinTRIV}, is of order $1-6\times 10^{-4}$.
$R^{PTRIV}_{em,2}$
for the $\beta^+$ decay process, which contains only second forbidden
retardation terms (namely, terms of order $\kappa R$ with $R$ some
typical radius of a nucleus), is of order or less than $5\times 10^{-6}$.
Figs. 24 and 26 are 3--dimensional plots of
$R^{PTRIV}_{em,1}$
and $R^{PTRIV}_{em,2}$ for the $\beta^+$ decay process of the A=8
system. $R^{PTRIV}_{em,1}$ for the $\beta^-$ decay process is
 smaller than the corresponding one in the $\beta^+$
decay process due to the destructive interference of $f_c$ and
$f_M$. It is of order or less than $10^{-3}$. $R^{PTRIV}_{em,2}$ for the
$\beta^-$ decay process depends only on the second forbidden
retardation terms. It is of order or less than $2\times 10^{-4}$.
 From these graphs, it can be seen that the observables proportional to
$\hat{\bf k}_\nu \hat{\bf
k}_e\times\hat{\bf k}_\nu$ in both $\beta^+$ and $\beta^-$ decay
processes contain smaller leptonic Coulomb interaction
induced PTRIV than other observables; thus it is advantageous to use
these observables to study the
question of genuine time reversal violation.
The $E_r$ dependence of $R_{em,i}^{PTRIV}$ (i=1,2) is determined by
the energy dependence of $\eta_6$, $\eta_8$,\ldots,$\eta_{10}$.
As we can see, the PTRIV signal due to
the $e^\pm$--nucleus
Coulomb interaction reaches its maximum in the energy region
$E_r\sim 3$ MeV, which
is in contrast to the hadronic PTRIV. Therefore, although the $e^\pm$--nucleus
 Coulomb interaction is suppressed by a factor of $Z\alpha$,
the magnitude of the PTRIV effects resulting from the hadronic strong
interaction are smaller than those due to the $e^\pm$--nucleus Coulomb
interaction in the energy region $E_r = 3$ MeV. Of course, the hadronic
strong (and Coulomb) PTRIV can be bigger if one goes away from the
the $E_r=3$ MeV region, in which different hadronic PTRIV amplitudes cancel
each other.

\section{Conclusions and Acknowledgments}
In this paper, we have examined a time reversal test in the $\beta$--decays of
the mass 8 nuclei.  An R--matrix treatment was used, with parameters related
to the $\alpha$--$\alpha$ scattering phase shifts and allowed $\beta$--decay
rates. The $e^\pm$--nucleus Coulomb final state interactions
are included.  False time--reversal violation (PTRIV) signals could arise due
to these $e^\pm$--nucleus Coulomb as well as the strong interaction effects in
the A=8 system. The strong interaction induced PTRIV is much smaller than the
$e^\pm$--nucleus Coulomb interaction ones because the former is  due to
several second forbidden contributing terms that nearly cancel each other in
the $E_r \approx$ 3--5 MeV region.  In the energy region close to the
lowest $2^+$ state around 3 MeV these contributions can be further minimized
because they have zeros.  The $e^\pm$--nucleus final state interaction effects
are $ \lsim 7 \times 10^{-4}$.  These effects have not been seen so far.

The authors are particularly grateful to Dr. Ludwig De Braeckeleer for many
useful discussions of the experiment and associated theory.  We also thank
Eric Adelberger for helpful comments.

%
%  end section
%
\setcounter{equation}{0}
%\input{refs1}
%
%  refs1.tex
%

%
%  end refs1
%
\setcounter{equation}{0}
%\input{appendic}
%
% appendic.tex
%
%
\setcounter{equation}{0}
\renewcommand{\theequation}{A.\arabic{equation}}
\setcounter{footnote}{0}
\renewcommand{\thefootnote}{A.\arabic{footnote}}
\section*{Appendix A: Hadronic Response Function in Terms of the
Reduced Matrix Elements of Multipole Operators}

Following Ref. \cite{thesis},
\be
R_{ss}(\sigma) &=& \sqrt{4\pi}\tilde\sigma
\sum  i^{J-J'}\ninj{J_f}{J'_f}{\sigma}{J_i}{J'_i}{0}
                        {J}{J'}{\sigma} F_{ss}(JJ';\sigma)
,\label{Rss}\\
R_{sv}^{(\pm)}(\sigma\rho) &=&\sqrt{4\pi}\tilde\sigma
\sum i^{J-J'}\ninj{J_f}{J'_f}{\sigma}{J_i}{J'_i}{0}{J}{J'}{\sigma}
F_{sv}^{(\pm)}(JJ';\sigma\rho),\label{Rsvpm}\\
R_{vv}(\sigma\rho\tau) &=& \sqrt{4\pi}
\tilde\sigma
\sum i^{J-J'}\ninj{J_f}{J'_f}{\sigma}{J_i}{J'_i}{0}{J}{J'}{\sigma}
F_{vv}(JJ';\sigma\rho\tau),\label{Rvv}
\ee
where $\hat x=\sqrt{2x+1}$, in order to simplify the notation, a
sum over indices not on the l.h.s. of the above equations is
implied.  It is found that
\be
F_{ss}(JJ';\sigma) &=& (-1)^J{\braa{\sigma}Y_{J'}\kett{J}
          \over\tilde\sigma} {\cal C}_J {\cal C}^*_{J'},\label{Fss}\\
F_{sv}^{\pm}(JJ';\sigma\rho)&=&
                  f_{sv}(JJ';\sigma\rho)\pm
                  f_{vs}(JJ';\sigma\rho),
\label{Fsvpm}\\
F_{vv}(JJ';\sigma\rho\tau)&=&(-1)^\rho\tilde J^2\tilde{J'}^2\tilde\tau
           \sum_{ll'}(-1)^{l+1}\braa{\rho}Y_{l'}\kett{l}\left\{
          \begin{array}{ccc}
               1&l'&J'\\
               1&l &J\\
              \tau&\rho&\sigma
          \end{array}\right\} a_l^J a_{l'}^{J'*},\nonumber\\ &&
\label{Fvv}
\ee
and
\be
f_{sv}(JJ';\sigma\rho) &=& (-1)^\sigma
           \tilde {J'}^2\sum_{l}\braa{\rho}Y_J\kett{l}
           \left \{\begin{array}{ccc}
               J&l&\rho\\
               1&\sigma&J'
          \end{array}\right \} {\cal C}_Ja_l^{J'*},
\label{fsv}\\
f_{vs}(JJ';\sigma\rho) &=& (-1)^{J+1}\tilde {J}^2\sum_l(-1)^l\braa{\rho}
           Y_{J'}\kett{l}
          \left\{ \begin{array}{ccc}
               J'&l&\rho\\
              1&\sigma&J
           \end{array}\right \} a_l^J {\cal C}_{J'}^*,
\label{fvs}
\ee
and
\be
a_{J-1}^J &=& (-1)^J{1\over 2J+1}\left (\sqrt{J+1} {\cal E}_J
+ \sqrt{J} {\cal L}_J\right ),
\label{ajm}\\
a_J^J &=& (-1)^J {\cal M}_J,
\label{aj0}\\
a_{J+1}^J &=& (-1)^J{1\over 2J+1}\left (\sqrt{J}{\cal E}_J -\sqrt{J+1}
{\cal L}_J\right ).
\ee
The reduced matrix elements ${\cal C}_J$, ${\cal L}_J$, ${\cal E}_J$ and
${\cal M}_J$ of the hadronic charged weak  currents can be decomposed into
\be
{\cal C}_J &=& C_J + C^5_J=\braa{J_f}\hat C_J\kett{J_i},\label{CCJ}\\
{\cal L}_J &=& L_J + L^5_J=\braa{J_f}\hat L_J\kett{J_i},\label{LLJ}\\
{\cal E}_J &=& E_J + E^5_J=\braa{J_f}\hat T^{el}_J\kett{J_i},\label{EEJ}\\
{\cal M}_J &=& M_J + M^5_J=\braa{J_f}\hat T^{mag}_J\kett{J_i},\label{MMJ}
\ee
where superscript ``5'' indicates that the corresponding matrix element is
originated from the axial vector current and the rest of the matrix elements
are originated from the vector current.

\setcounter{equation}{0}
\renewcommand{\theequation}{B.\arabic{equation}}
\setcounter{footnote}{0}
\renewcommand{\thefootnote}{B.\arabic{footnote}}
\section*{Appendix B: Power Expansion of the Hadronic Response Functions
 for $A=8$ System in Terms of Momentum Transfer}

  The relation between $R_1\ldots R_{12}$ and $R_{ss}\ldots R_{vv}$ can be
found. For T--even observables, only the final $2^+$ state contributions
are considered. While for T--odd observables, we consider contributions
from the final $2^+$ state, and from the interference between the final
$2^+$ state with the final $0^+$ and $4^+$ states. In the following discussion,
we shall introduce set of response functions $\bar R_{ss},\ldots,\bar R_{vv}$
that are related to $R_{ss},\ldots,R_{vv}$ through
\be
\bar R_i &=& {R_i\over |g_A|^2 |A_2|^2},\hskip 1 in (i=ss,\ldots,vv)
\label{Rbar}
\ee
where $A_2$ is the Gamow--Teller matrix element defined in Table
\ref{StaticMop}. With
these set of response functions, the Cartesian ones can be found.
The mass of the charged leptons $m_e$ can be neglected; we do so in the
following.
Those correspond to
$\sigma=0$ and T--even observables are
\be
R_1^{(0)}&=& 5\bar R_{ss}(0) + 5{\Delta\over\kappa}\bar R_{sv}^{(+)}(01)
                -5\sqrt{3}\bar R_{vv}(000)\pm 5\sqrt{2}{\epsilon_e-\epsilon_\nu
                   \over \kappa}\bar R_{vv}(011)\nonumber\\
          && + {50\over 3}\sqrt{2\over
                   3}{\epsilon_e\epsilon_\nu\over\kappa^2}\bar R_{vv}(022),
\label{R10}\\
R_2^{(0)}&=& 5\bar R_{ss}(0) + 5{\Delta\over\kappa}\bar R_{sv}^{(+)}(01)
                 +{5\over\sqrt{3}}\bar R_{vv}(000)\mp 5\sqrt{2}
               {\epsilon_e-\epsilon_\nu\over\kappa}\bar R_{vv}(011)\nonumber\\
          && + 10\sqrt{2\over 3} {\epsilon^2_e+\epsilon^2_\nu\over\kappa^2}
              \bar R_{vv}(022),
\label{R20}\\
R_3^{(0)}&=& 5\sqrt{2\over 3}{\epsilon_e\epsilon_\nu\over\kappa^2}
             \bar  R_{vv}(022).
\label{R30}
\ee
Those correspond to $\sigma=2$, T--even, which contain only the final
$2^+$ state contributions, are
\be
R_1^{(2)}&=& -{15\over\sqrt{7}}{\eps_e^2\over\kappa^2}\bar R_{ss}(2) +
             {15\over\sqrt{7}}{\eps_e\over\kappa}\bar R_{sv}^{(+)}(21)\mp
            5\sqrt{15\over 7}{\eps_e\eps_\nu\over\kappa^2}\bar R_{sv}^{(-)}(22)
             \pm 15\sqrt{2\over 7}{\eps_e\over\kappa}\bar R_{vv}(211)
            \nonumber\\
            &&
          + {75\over 2}\sqrt{6\over 35}{\eps_e^2\over\kappa^2}\bar R_{vv}(220)
            +{15\over 7}\sqrt{30}{\eps_e\eps_\nu\over\kappa^2}\bar R_{vv}(222),
\label{R12}\\
R_2^{(2)}&=& -{30\over\sqrt{7}}{\eps_e\eps_\nu\over\kappa^2}\bar R_{ss}(2)
            +{15\over\sqrt{7}}{\Delta\over\kappa}\bar R_{sv}^{(+)}(21)
         \pm 5\sqrt{15\over 7}{\eps^2_e-\eps^2_\nu\over\kappa^2}
                     \bar R_{sv}^{(-)}(22)
         \nonumber\\
        && -10\sqrt{3\over 7}\bar R_{vv}(202)\mp 15\sqrt{2\over
            7}{\eps_e-\eps_\nu\over\kappa}\bar R_{vv}(211)+75\sqrt{6\over 35}
          {\eps_e\eps_\nu\over\kappa^2}\bar R_{vv}(220)
         \nonumber\\
        &&+{5\over 7}\sqrt{30}{\eps_e^2+\eps^2_\nu\over\kappa^2}\bar
R_{vv}(222),
\label{R22}\\
R_3^{(2)}&=& -{15\over\sqrt{7}}{\eps_\nu^2\over\kappa^2}\bar R_{ss}(2)
            +{15\over\sqrt{7}}{\eps_\nu\over\kappa}\bar R_{sv}^{(+)}(21)
           \pm 5\sqrt{15\over 7}{\eps_e\eps_\nu\over\kappa^2}
           \bar R_{sv}^{(-)}(22)
          \nonumber\\
        &&\mp 15\sqrt{2\over 7}{\eps_\nu\over\kappa}\bar R_{vv}(211)
          +{75\over 2}\sqrt{6\over 35}{\eps_\nu^2\over\kappa^2}\bar R_{vv}(220)
          +{15\over 7}\sqrt{30}{\eps_e\eps_\nu\over\kappa^2}\bar R_{vv}(222),
\label{R32}\\
R_4^{(2)}&=&-{15\over\sqrt{7}}{\eps_e^2\over\kappa^2}\bar R_{ss}(2)\mp
             5\sqrt{15\over 7} {\eps_e^2\over\kappa^2}\bar R_{sv}^{(-)}(22)
            -{25\over 2}\sqrt{6\over 35}{\eps^2_e\over\kappa^2}\bar R_{vv}(220)
           \nonumber\\
          && +{5\over 7}\sqrt{30}{\eps^2_e\over\kappa^2}\bar R_{vv}(222),
\label{R42}\\
R_5^{(2)}&=&-{30\over\sqrt{7}}{\eps_e\eps_\nu\over\kappa^2}\bar R_{ss}(2)
            -25\sqrt{6\over 35}{\eps_e\eps_\nu\over\kappa^2}\bar R_{vv}(220)
            -{10\over 7}\sqrt{30}{\eps_e\eps_\nu\over\kappa^2}\bar R_{vv}(222),
\label{R52}\\
R_6^{(2)}&=&-{15\over\sqrt{7}}{\eps_\nu^2\over\kappa^2}\bar R_{ss}(2)
           \pm 5\sqrt{15\over 7}{\eps_\nu^2\over\kappa^2}\bar R_{sv}^{(-)}(22)
           -{25\over 2}\sqrt{6\over 35}{\eps_\nu^2\over\kappa^2}\bar
R_{vv}(220)
\nonumber\\
          && -{5\over 7}\sqrt{30}{\eps_\nu^2\over\kappa^2}\bar R_{vv}(222).
\label{R62}
\ee
Those correspond to $\sigma=2$, T--odd, which contain only the final
$2^+$ state contributions are
\be
R_7^{(2)}&=& 15\sqrt{5\over 21}{\eps_e(\eps_e-\eps_\nu)\over\kappa^2}
             Im\bar R_{sv}^{(+)}(22)\pm {15\over\sqrt{7}}{\eps_e\over\kappa}
             Im\bar R_{sv}^{(-)}(21)\nonumber\\ &&\mp 5\sqrt{30\over 7}
             {\eps_e(\eps_e-\eps_\nu)\over\kappa^2}Im\bar R_{vv}(221)
             -5\sqrt{6\over 7}{\eps_e\over\kappa}Im\bar R_{vv}(212),
\label{R72}\\
R_8^{(2)}&=&15\sqrt{5\over 21}{\eps_\nu(\eps_e-\eps_\nu)\over\kappa^2}
             Im\bar R_{sv}^{(+)}(22)\pm {15\over\sqrt{7}}{\eps_\nu\over\kappa}
             Im\bar R_{sv}^{(-)}(21)\nonumber\\ &&\mp 5\sqrt{30\over 7}
             {\eps_\nu(\eps_e-\eps_\nu)\over\kappa^2}Im\bar R_{vv}(221)
             +5\sqrt{6\over 7}{\eps_\nu\over\kappa}Im\bar R_{vv}(212).
\label{R82}
\ee
Those correspond to $\sigma=2$, T--odd, which originate from the
interference contributions between final $2^+$ state and final
$0^+$ state, are
\be
R_9^{(2)}&=& sin(\delta_2-\delta_0)\left [
             15\sqrt{2\over 3} {\eps_e(\eps_e-\eps_\nu)\over\kappa^2}
             Re\left (\bar R_{sv}^{(+)}(22)\right)_{20}\pm 15\sqrt{2\over 5}
             {\eps_e\over\kappa}Re\left(\bar
R_{sv}^{(-)}(21)\right)_{20}\right.
\nonumber\\
          &&\left.\mp 10\sqrt{3}{\eps_e(\eps_e-\eps_\nu)\over\kappa^2}
             Re\left(\bar
             R_{vv}(221)\right)_{20} - 10\sqrt{3\over 5}{\eps_e
             \over\kappa} Re\left (\bar R_{vv}(212)\right )_{20}\right]
\nonumber\\
          &&  -cos(\delta_2-\delta_0)\left [
             15\sqrt{2\over 3} {\eps_e(\eps_e-\eps_\nu)\over\kappa^2}
             Im\left (\bar R_{sv}^{(+)}(22)\right)_{20}\pm 15\sqrt{2\over 5}
             {\eps_e\over\kappa}Im\left(\bar
R_{sv}^{(-)}(21)\right)_{20}\right.
\nonumber\\
          &&\left.\mp 10\sqrt{3}{\eps_e(\eps_e-\eps_\nu)\over\kappa^2}
             Im\left(\bar R_{vv}(221)\right)_{20}-10\sqrt{3\over 5}{\eps_e
             \over\kappa} Im\left (\bar R_{vv}(212)\right )_{20}]
\right ],
\label{R92}\\
R_{10}^{(2)}&=& sin(\delta_2-\delta_0)\left [
             15\sqrt{2\over 3} {\eps_\nu(\eps_e-\eps_\nu)\over\kappa^2}
             Re\left (\bar R_{sv}^{(+)}(22)\right)_{20}\pm 15\sqrt{2\over 5}
             {\eps_\nu\over\kappa}
             Re\left(\bar R_{sv}^{(-)}(21)\right)_{20}\right.
\nonumber\\
          &&\left.\mp 10\sqrt{3}{\eps_\nu(\eps_e-\eps_\nu)\over\kappa^2}
             Re\left(\bar
             R_{vv}(221)\right)_{20} + 10\sqrt{3\over 5}{\eps_\nu
             \over\kappa} Re\left (\bar R_{vv}(212)\right )_{20}\right ]
\nonumber\\
          &&  -cos(\delta_2-\delta_0)\left [
             15\sqrt{2\over 3} {\eps_\nu(\eps_e-\eps_\nu)\over\kappa^2}
             Im\left (\bar R_{sv}^{(+)}(22)\right)_{20}\pm 15\sqrt{2\over 5}
             {\eps_\nu\over\kappa}
             Im\left(\bar R_{sv}^{(-)}(21)\right)_{20}\right.
\nonumber\\
          &&\left.\mp 10\sqrt{3}{\eps_\nu(\eps_e-\eps_\nu)\over\kappa^2}
             Im\left(\bar R_{vv}(221)\right)_{20}+10\sqrt{3\over 5}{\eps_\nu
             \over\kappa} Im\left (\bar R_{vv}(212)\right )_{20}]
\right ].\nonumber\\&&\nonumber\\ &&
\label{R102}
\ee
Those correspond to $\sigma=2$, T--odd, which originate from the
interference contributions between final $2^+$ state and final
$4^+$ state, are
\be
R_{11}^{(2)} &=& sin(\delta_2-\delta_4)\left [
              30\sqrt{3\over 7}{\eps_e(\eps_e-\eps_\nu)\over\kappa^2}
             Re\left(\bar R_{sv}^{(+)}(22)\right )_{24}\pm
             18\sqrt{5\over 7}{\eps_e\over\kappa}
              Re\left(\bar R_{sv}^{(-)}(21)\right)_{24}\right.
\nonumber\\&&\left.
             \mp 30\sqrt{6\over 7}{\eps_e(\eps_e-\eps_\nu)\over\kappa^2}
            Re\left (\bar R_{vv}(221)\right)_{24}-6\sqrt{30\over 7}
            {\eps_e\over \kappa}Re\left (\bar R_{vv}(212)\right )_{24}\right ]
\nonumber\\
           &&   -cos(\delta_2-\delta_4)\left [
              30\sqrt{3\over 7}{\eps_e(\eps_e-\eps_\nu)\over\kappa^2}
             Im\left(\bar R_{sv}^{(+)}(22)\right )_{24}\pm
             18\sqrt{5\over 7}{\eps_e\over\kappa}
              Im\left(\bar R_{sv}^{(-)}(21)\right)_{24}\right.
\nonumber\\&&\left.
             \mp 30\sqrt{6\over 7}{\eps_e(\eps_e-\eps_\nu)\over\kappa^2}
            Im\left (\bar R_{vv}(221)\right)_{24}-6\sqrt{30\over 7}
            {\eps_e\over \kappa}Im\left (\bar R_{vv}(212)\right )_{24}\right ],
\label{R112}\\
R_{12}^{(2)} &=& sin(\delta_2-\delta_4)\left [
              30\sqrt{3\over 7}{\eps_\nu(\eps_e-\eps_\nu)\over\kappa^2}
             Re\left(\bar R_{sv}^{(+)}(22)\right )_{24}\pm
             18\sqrt{5\over 7}{\eps_\nu\over\kappa}
              Re\left(\bar R_{sv}^{(-)}(21)\right)_{24}\right.
\nonumber\\&&\left.
             \mp 30\sqrt{6\over 7}{\eps_\nu(\eps_e-\eps_\nu)\over\kappa^2}
            Re\left (\bar R_{vv}(221)\right)_{24}+6\sqrt{30\over 7}
            {\eps_\nu\over \kappa}Re\left (\bar R_{vv}(212)\right )_{24}\right
]
\nonumber\\
           &&   -cos(\delta_2-\delta_4)\left [
              30\sqrt{3\over 7}{\eps_\nu(\eps_e-\eps_\nu)\over\kappa^2}
             Im\left(\bar R_{sv}^{(+)}(22)\right )_{24}\pm
             18\sqrt{5\over 7}{\eps_\nu\over\kappa}
              Im\left(\bar R_{sv}^{(-)}(21)\right)_{24}\right.
\nonumber\\&&\left.
             \mp 30\sqrt{6\over 7}{\eps_\nu(\eps_e-\eps_\nu)\over\kappa^2}
            Im\left (\bar R_{vv}(221)\right)_{24}+6\sqrt{30\over 7}
            {\eps_\nu\over \kappa}
            Im\left (\bar R_{vv}(212)\right )_{24}\right ],
\label{R122}
\ee
The subindices ``20'' and ``24'' in Eqs. \ref{R92}--\ref{R122} denote they are
the interference terms between the $2^+$ state and $0^+$ or $4^+$ respectively.
Response function $R_{ss},\ldots,R_{vv}$ are related to the reduced matrix
elements of the multipole operators of the hadronic charged weak currents;
they are given in Appendix A.
Using the definition for the static multipoles given in Tables
\ref{MultOpExpand} and \ref{StaticMop}, the
response functions are ready to be expressed in terms of $f_1$, $f_M$, $f_T$,
$f^5_c$
and $\eta_k$ ($k=1,\ldots,10$). (See Eqs. 2.41--2.45.)
The resulting T--even response functions have
the following form
\be
R_1^{(0)} &=& 1 + {\Delta\over 3M} Re\left
          (f_c^5\mp 2f_M\mp 2f_T-\sqrt{3\over 2}
             {\Delta\over M}\eta_7\right )\nonumber\\
           &&+{4\eps_e\over 3M}Re\left (\pm f_M+{5\over 6}\sqrt{2\over
           3}{\Delta\over M}\eta_7 + {\sqrt{2}\over 9}{\Delta\over M}\eta_8
          \right )\nonumber\\
           && -{2\eps_e^2\over 9M^2}Re\left ( 5\sqrt{2\over 3}\eta_7
               + {2\sqrt{2}\over 3}\eta_8\right ),
\label{Rte1}\\
R_2^{(0)} &=& -{1\over 3} + {\Delta\over 3M}Re\left ( f_c^5\pm 2f_M\pm 2f_T
              +{1\over 3}\sqrt{3\over 2}{\Delta\over M}\eta_7+{2\sqrt{2}\over
               9}{\Delta\over M}\eta_8\right)\nonumber\\
           &&  + {4\eps_e\over 3M}Re\left (\mp f_M - \sqrt{2\over 3}
               {\Delta\over M}\eta_7 -{2\sqrt{2}\over 15}{\Delta\over M}
              \eta_8\right )\nonumber\\
            && +{4\eps_e^2\over 9M^2}Re\left (\sqrt{6}\eta_7
              + {2\sqrt{2}\over 5}\eta_8\right ),
\label{Rte2}\\
R_3^{(0)} &=& {\Delta\eps_e\over 9M^2}Re\left (\sqrt{6}\eta_7 + {2\sqrt
2\over5}
               \eta_8\right )-{\eps_e^2\over 9M^2}Re\left (\sqrt{6}\eta_7
               +{2\sqrt{2}\over 5}\eta_8\right ),
\label{Rte3}\\
R_1^{(2)} &=& {\eps_e\over 2M}Re\left (-f_c^5\pm f_M\mp f_T\mp {1\over 3}
             \sqrt{3\over 35}{\Delta\over M}f_1\eta_6
             -{\sqrt{2}\over 5}{\Delta\over M}\eta_8
             + {1\over 5}\sqrt{10\over 7}{\Delta\over M}\eta_9\right.
\nonumber\\ &&\left.
             +{8\over 35}{1\over\sqrt{7}}{\Delta\over M}\eta_{10}
              \right )
             + {\eps_e^2\over M^2}Re\left (\mp {1\over 7}\sqrt{7\over 15}f_1
              \eta_6
              -{1\over 5}\sqrt{10\over 7}\eta_9 + {2\over
              7\sqrt{7}}\eta_{10}\right ),
\label{Rte4}\\
R_2^{(2)} &=& -1 + {\Delta\over 2M}Re\left (-f_c^5 \pm f_M\pm f_T\mp{1\over 3}
              \sqrt{3\over
               35}{\Delta\over M}f_1\eta_6 +{2\Delta\over 3M}\eta_7
               - {\sqrt{2}\over 15}{\Delta\over M}
               \eta_8\right.\nonumber \\
          && \left. +{1\over 15}\sqrt{10\over 7}{\Delta\over M}\eta_9
              +{8\over 105\sqrt{7}}{\Delta\over M}\eta_{10}
            \right ) + {\eps_e\over M}Re\left (
            \mp f_M \pm {1\over 3}\sqrt{3\over 35}{\Delta\over M} f_1\eta_6
            \right.\nonumber\\
          &&\left. -{2\Delta\over 3M}\eta_7-{2\sqrt{2}\over 15}{\Delta\over M}
            \eta_8
           -{4\over 3}\sqrt{2\over 35}{\Delta\over M}\eta_9
             +{76\over 105\sqrt{7}}{\Delta\over M}\eta_{10}\right )\nonumber\\
          && +{\eps^2_e\over M^2}Re\left ({2\over 3}\eta_7 + {2\sqrt{2}\over
             15}\eta_8 + {4\over 3}
            \sqrt{2\over 35}\eta_9  - {76\over 105\sqrt{7}}\eta_{10}
            \right ),
\label{Rte5}\\
R_3^{(2)} &=& {\Delta\over 2M}\left (-f_c^5\mp f_M\mp f_T\pm
               \sqrt{3\over 35}{\Delta\over M}f_1\eta_6
               -{\sqrt 2\over
                5}{\Delta\over M}\eta_8-\sqrt{2\over 35}{\Delta\over M}\eta_9
               + {4\over 5\sqrt{7}}{\Delta\over M}
               \eta_{10}\right ) \nonumber\\
           &&
               + {\eps_e\over 2M}Re\left (
               f_c^5\pm f_M\pm f_T\mp \sqrt{5\over 21}{\Delta\over M}
               f_1\eta_6 +{\sqrt 2\over 5}{\Delta\over M}\eta_8
             + 3\sqrt{2\over 35}{\Delta\over M}\eta_9
		\right.\nonumber \\
            && \left.
               - {48\over 35\sqrt{7}}
              {\Delta\over M}\eta_{10}\right )
              +{(\Delta-\eps_e)^2\over M^2}
              Re\left (\pm {1\over 7}\sqrt{7\over 15}
              f_1\eta_6
              -\sqrt{2\over 35}\eta_9 +
              {2\over 7\sqrt{7}}\eta_{10}\right ),
\nonumber\\ &&
\label{Rte6}\\
R_4^{(2)} &=& {\eps_e^2\over M^2}Re\left (
             \pm {1\over 7}\sqrt{7\over 15} f_1\eta_6 +
             {1\over 15}\sqrt{10\over 7}\eta_9 -
             {2\over 21\sqrt{7}}\eta_{10}\right ),
\label{Rte7}\\
R_5^{(2)} &=& {\eps_e\over M}Re\left ({2\over 3}{\Delta\over M}\eta_7
             + {2\sqrt{2}\over 15}{\Delta\over M}\eta_8
             -{12\over 35\sqrt{7}}
            {\Delta\over M}\eta_{10}\right )\nonumber\\
          && +{\eps_e^2\over M^2}Re
            \left ( -{2\over 3}\eta_7 -{2\sqrt{2}\over 15}
            \eta_8 + {12\over 35\sqrt{7}}
           \eta_{10}\right ),
\label{Rte8}\\
R_6^{(2)} &=& {\Delta^2\over M^2}Re\left (\mp {1\over 7}\sqrt{7\over 15}
              f_1\eta_6 - {\sqrt{2}\over 15}\eta_8
              + {6\over 35\sqrt{7}}
             \eta_{10}\right )\nonumber\\
          && +{\Delta\eps_e\over 3M^2}
            Re\left (\pm {2\over 7}\sqrt{7\over 15}f_1\eta_6
            -{2\sqrt{2}\over 5}\eta_8
            +{12\over 35\sqrt{7}}\eta_{10}\right )\nonumber\\
          &&+{\eps_e^2\over M^2}Re\left (\mp {1\over 7}\sqrt{7\over
15}f_1\eta_6
             +{\sqrt{2}\over 15}\eta_8 -
                {6\over 35\sqrt{7}}\eta_{10}\right ).
\label{Rte9}
\ee
Before ending this appendix, some more details are worth mentioning.
The general
form of response functions Eqs. \ref{R10}--\ref{R62} depend on ${\bf k}_e\cdot
{\bf k}_\nu$ though their dependence on $\kappa$. When the power expansion in
terms of $\kappa$ to order $O(\kappa^2)$ is performed, these dependence are
extracted to make the response function defined in
Eqs. \ref{Rte1}--\ref{Rte9} independent of ${\bf k}_e\cdot{\bf k}_\nu$.
Therefore $R_i^{(0)}$ (i=1,2,3) and $R_i^{(2)}$ (i=1,\ldots,6) defined in
Eqs. \ref{R10}--\ref{R62} are slightly different from those defined in
Eqs. \ref{Rte1}--\ref{Rte9}, which are independent of ${\bf k}_e\cdot {\bf
k}_\nu$. However, the differences are small and only depend on $\eta_7$.

\setcounter{equation}{0}
\renewcommand{\theequation}{C.\arabic{equation}}
\setcounter{footnote}{0}
\renewcommand{\thefootnote}{C.\arabic{footnote}}
\section*{Appendix C: Non--relativistic Reduction of the One Nucleon Matrix
Elements of the Hadronic Weak Current Operators }

\be
\rho^{(\pm)}({\bf x}) & = & F_1^V\sum_{i=1}^A \tau^{(\pm)}(i)\delta({\bf x}-
{\bf r}_i),\label{rhopm}\\
\rho^{(\pm)5}({\bf x}) &=& \mp {ig_T\over 2M}\sum_{i=1}^A \tau^{(\pm)}(i)
\sigma(i)\cdot\nabla\delta({\bf x}-{\bf r}_i)\nonumber\\
&& + g_A\sum_{i=1}^A\tau^{(\pm)}(i)\sigma(i)\cdot\left [
{{\bf p}_i\over 2M}\delta({\bf x}-{\bf r}_i) + \delta({\bf x}-{\bf r}_i){{\bf
p}_i\over 2M}\right ],\label{rho5pm}\\
J^{(\pm)}({\bf x}) &=& F_1^V\sum_{i=1}^A\tau^{(\pm)}(i)\left
[{{\bf p}_i\over 2M}\delta({\bf x}-{\bf r}_i) + \delta({\bf x}-{\bf r}_i){{\bf
p}_i\over 2M}\right ]\nonumber\\
&& + {G^V_M\over 2M}\sum_{i=1}^A \tau^{(\pm)}(i)\nabla\times\sigma(i)
\delta({\bf x}-{\bf r}_i),\label{tJpm}\\
A^{(\pm)}({\bf x}) &=& \left (g_A\pm{\Delta\over 2M} g_T\right )\sum_{i=1}^A
\tau^{(\pm)}(i)\sigma(i)\delta({\bf x}-{\bf r}_i),\label{tApm}
\ee
where $\Delta=E_i-E_f$ is the difference of the initial and the final
energies of the hadronic system, $G^V_M = F_M^V + 1 = 4.71$, terms
that depend
on $g_P$ are not included since their contribution to the differential
decay rate is extremely small when the charged lepton polarizations
are not detected \cite{SYing}.

\setcounter{equation}{0}
\renewcommand{\theequation}{D.\arabic{equation}}
\setcounter{footnote}{0}
\renewcommand{\thefootnote}{D.\arabic{footnote}}
\section*{Appendix D: R--matrix theory for the $\beta$-decay processes
  of the A = 8 system}

We only present an outline of the theory here; for more details we refer the
reader to
Refs. [19,26].

To the first order in the weak interaction Hamiltonian $H_w$ and
all orders in the strong and electromagnetic interaction Hamiltonian
$H_s$ and $H_{em}$, the $S$--matrix for
the $^8$B or $^8 Li$ $\to$ $2\alpha$ transition can be written as
\be
S_{fi} &=& -i \int_{-\infty}^{\infty} dt\bra{2\alpha;e\nu}U(\infty,t)
H_w U(t,-\infty)\ket{\phi_i},\label{Smatrix41}
\ee
where $U(t_1,t_2)$ is the full propagator of the system under strong and
electromagnetic interactions
and $\ket{\phi_i}$ is either $^8$B or $^8 Li$.
With a proper phase convention for $\ket{\phi_i}$, which is regarded as
an eigenstate of $H=T+H_s+H_{em}$ (T is the kinetic energy operator),
\be
S_{fi} &=& -i \int_{-\infty}^{\infty} dt e^{-iE_i t}\bra{2\alpha;e\nu}
U(\infty,t)H_w \ket{\phi_i}.\label{Smatrix42}
\ee
A complete set of states can be inserted between the propagator and the
weak interaction Hamiltonian $H_w$,
\be
S_{fi} &=& -i\int_{-\infty}^{\infty}dt e^{-iE_i t}
       \sum_I \bra{2\alpha;e\nu}
       U(\infty,t)\ket{I}\bra{I}H_w\ket{\phi_i},
\label{Smatrix43}
\ee
which is illustrated in Fig. 1.
Since the matrix elements of the
weak interaction Hamiltonian $H_w$ are large only when the hadronic part of
the state $\ket{I}$
is comparable in spatial extent to that of $\ket{\phi_i}$
\footnote{It is possible that some states in the continuum
also have large matrix elements. In case of the A=8 system, it does not
turn out to be the case. It will be demonstrated in the following that
the direct transition from $^8$B or $^8$Li to two $\alpha$ particle
scattering states characterized by a parameter C is small.},
the summation in Eq. \ref{Smatrix43}
can be saturated to a satisfactory precision by a subset of states with
comparable spatial extensions to $\ket{\phi_i}$. These sets of states can
be generated by choosing
eigenstates of a Hamiltonian that confine them in a region with a size that
is comparable to $\ket{\phi_i}$. It is natural
to choose the direct product  of
shell model states $\ket{n}$ and
the leptonic states $\ket{e\nu}$. The shell model Hamiltonian is so constructed
that the energies  of its eigenstates are approximately located at the
resonant peaks\footnote{i.e.
the energies at which the phase shifts of the $\alpha$--$\alpha$ scattering
S--matrix pass through
$(n+{1\over 2})\pi$ from below, with $n=0,\pm 1,\ldots$.} of the $^8$Be$^*$
system.
At low energies,
only a very few such shell model states are important,
since the higher excitations of
the shell model Hamiltonian have energies well separated from the
ground state.

The strong and electromagnetic interaction Hamiltonian can be written as
\be
H_s+H_{em} &=& H_{shell} + \Delta V,\label{Hshell}
\ee
where $H_{shell}$ is the shell model Hamiltonian and $\Delta V$ is
the residual interaction that is responsible for the decay of the
shell model $^8$Be$^*$ states. The decay of the localized state
$\ket{n}$, of interest in this paper, to an asymptotic two $\alpha$
state can be parameterized in terms of the $R$--matrix \cite{GWscat} as
\be
R_{2\alpha,n}(t) &=& \bra{2\alpha;e\nu}U(\infty,t)\ket{n},\nonumber\\
&=&\int_{C}{dE\over 2\pi i} e^{iE t}{1\over E-E_{2\alpha}}
\Gamma_{2\alpha,n}(E){1\over E-E_n-R_n(E)},\label{Rdef1}
\ee
where the path of the integration in the complex E plane is given in
Fig. 4.
The energy dependent function $\Gamma_{2\alpha,n}(E)$ is
the vertex function for
the decay of the shell model states $\ket{n}$ into the two $\alpha$ particle
states and the energy dependent function $R_n(E)$ is the self energy of the
state $\ket{n}$. They are all related to $\Delta V$ in the following
way
\be
R_n(E) &=& \bra{n} \Delta VF\ket{n},\label{RnE}\\
\Gamma_{2\alpha,n}(E) &=& \bra{2\alpha}\Delta VF\ket{n},\label{Gammaalphan}
\ee
where the state $F\ket{n} =\ket{\psi_n}$ satisfies a Lippmann--Schwinger
type of equation
\be
\ket{\psi^{\pm}_n} &=&\ket{n} + {1\over E-H_{shell}\pm i\epsilon}
                      \Delta V\ket{\psi_n^{\pm}},\label{LSeqforFn}
\ee
with $\ket{n}$ an eigenstate of $H_{shell}$.

%
% end appendic
%
%\input{fig}
%
%  fig.tex
%
\newpage
\noindent
{\bf Figure Captions}
\vspace{.5in}

\noindent
Figure 1: Diagrammatic representation of the $\beta$--decay in A=8 system.
\vspace{.25in}

\noindent
Figure 2: $^8Be$ energy levels calculated in the shell model compared with
the experimental ones. The shell model energy level at 13.37 MeV should be
identified with the experimental one at 16.92 MeV. It is explained in the
text. The shell model energy level at 16.19 MeV should be identified with
the experimental one at 16.63 MeV.
\vspace{.25in}

\noindent
Figure 3:  Experimental resonant levels in the $^8Be^*$ system with $J^P=0^+$,
$2^+$ and $4^+$. The first column is the $0^+$ series, the second
column is the $2^+$ series and the third column is the $4^+$ series.
\vspace{.25in}

\noindent
Figure 4:  Result of the least square fit of the R--matrix parameters to the
         experimental phase shifts
         of the $J^P=0^+$ state.
\vspace{.25in}

\noindent
Figure 5:  Result of the least square fit of the R--matrix parameters to the
         experimental
         phase shifts of the $J^P=2^+$ state.
\vspace{.25in}

\noindent
Figure 6:  Result of the least square fit of the R--matrix parameters to the
         experimental
         phase shifts of the $J^P=4^+$ state.
\vspace{.25in}

\noindent
Figure 7:  R--matrix calculation of the $\alpha$ particle spectrum
in the $\beta$--decay process $^8B\to\alpha+\alpha+e^++\nu_e$. Solid
curve represents the full calculation and dashed curve represents
the calculation without the final potential scattering between
the final two $\alpha$ particles. Open circles represent the experimental
measured values. The contributions of the 3 MeV state and the 16 MeV
doublet states to the total counts are also indicated.

\vspace{.25in}

\noindent
Figure 8:  R--matrix calculation of the $\alpha$ particle spectrum
in the $\beta$--decay process $^8Li\to\alpha+\alpha+e^-+\bar\nu_e$.
Others are the same as Fig. 8.

\vspace{.25in}

\noindent
Figure 9:  R--matrix calculation of the $\alpha$ particle spectrum
in the $\beta$--decay process $^8B\to\alpha+\alpha+e^++\nu_e$. A linear
scale is used. Open circles represent the experimental measured
values.
\vspace{.25in}

\noindent
Figure 10: R--matrix calculation of the $\alpha$ particle spectrum
in the $\beta$--decay process $^8Li\to\alpha+\alpha+e^-+\bar\nu_e $.
Others are the same as Fig. 10.

\vspace{.25in}

\noindent
Figure 11:  Energy dependence of $\eta_6(2^+,0^+)$.
\vspace{.25in}

\noindent
Figure 12:  Energy dependence of $\eta_9(2^+,0^+)$.
\vspace{.25in}

\noindent
Figure 13: Energy dependence of $\eta_6(2^+,4^+)$.
\vspace{.25in}

\noindent
Figure 14: Energy dependence of $\eta_9(2^+,4^+)$.
\vspace{.25in}

\noindent
Figure 15: 3--dimensional plot of $R_1^{PTRIV}(E_r,\epsilon_e)$ for
the $\beta^+$ decay of A=8 system.  The vertical scale is 10$^{-5}$.

\vspace{.25in}

\noindent
Figure 16: 3--dimensional plot of $R_2^{PTRIV}(E_r,\epsilon_e)$ for
the $\beta^+$ decay of A=8 system.  The vertical scale is 10$^{-5}$
\vspace{.25in}

\noindent
Figure 17: 3--dimensional plot of $R_1^{PTRIV}(E_r,\epsilon_e)$ for
the $\beta^-$ decay of A=8 system.  The vertical scale is 10$^{-5}$.
\vspace{.25in}

\noindent
Figure 18:  3--dimensional plot of $R_2^{PTRIV}(E_r,\epsilon_e)$ for
the $\beta^-$ decay of A=8 system.  The vertical scale is 10$^{-5}$.
\vspace{.25in}

\noindent
Figure 19: Averaged PTRV response function $\bar R_1^{PTRV}(E_r)$ as an
function of $E_r$. Solid lines represent $\bar R_9^{(2)}+\bar R_{11}^{(2)}$.
Dashed lines represent $\bar R_9^{(2)}$ and $\bar R_{11}^{(2)}$ for the
$\beta^-$ decay. Dash--dotted line represent $\bar R_9^{(2)}$ and
$\bar R_{11}^{(2)}$ for the $\beta^+$ decay.
\vspace{.25in}

\noindent
Figure 20: Averaged PTRV response function $\bar R_2^{PTRV}(E_r)$ as an
function of $E_r$. Solid lines represent $\bar R_{10}^{(2)}+\bar R_{12}^{(2)}$.
Dashed lines represent $\bar R_{10}^{(2)}$ and $\bar R_{12}^{(2)}$ for the
$\beta^-$ decay. Dash--dotted line represent $\bar R_{10}^{(2)}$ and
$\bar R_{12}^{(2)}$ for the $\beta^+$ decay.
\vspace{.25in}

\noindent
Figure 21: 3--dimensional plot of $R^{PTRV}_{em,1}(E_r,\eps_e)$ for the
        $\beta^+$ decay of A=8 system.  The vertical scale is 10$^{-4}$.
\vspace{.25in}

\noindent
Figure 22: 3--dimensional plot of $R^{PTRV}_{em,1}(E_r,\eps_e)$ for the
$\beta^-$ decay of A=8 system.  The vertical scale is 10$^{-4}$.
\vspace{.25in}

\noindent
Figure 23:  3--dimensional plot of $R^{PTRV}_{em,2}(E_r,\eps_e)$ for the
        $\beta^+$ decay of A=8 system.  The vertical scale is 10$^{-5}$.
\vspace{.25in}

\noindent
Figure 24: 3--dimensional plot of $R^{PTRV}_{em,2}(E_r,\eps_e)$ for the
$\beta^-$ decay of A=8 system.  The vertical scale is 10$^{-4}$.
\vspace{.25in}

\noindent
Figure 25:  The path for the complex E integration. Solid dots represent
poles of the integrand and the dashed line represents the cut of the
integrand. (It is usually located on the real axis extending all the way to
positive infinity. Two cuts along different directions are equivalent provided
that there are no singularities in between them.)
\vspace{.25in}

\noindent
Figure 26: Diagrammatic representation of the resonant level contribution to
the $\alpha$--$\alpha$ scattering amplitude.
%
%  end fig
%
%\input{tab}
%
%  tab.tex
%

\newpage

\begin{table}[h]
\caption{Kinematic functions considered. Here ``T'' and ``P'' denote
  the time reversal and parity reflection transformations. \label{KineFunc}}
\vskip 0.2 in
\footnotesize{
\begin{tabular}{|c|c|c|}\hline\hline
\multicolumn{3}{|c|} {$\sigma=0$}\\ \hline
\head&Even&Odd\\
\hline
Even&\begin{tabular}{c}$K_{ss}(0)$ $K_{sv}^{(+)}(01)$ $K_{vv}(000)$

     $K_{vv}(011)$\\
     $K_{vv}(022)$
     \end{tabular} & \rlap{\hskip 1.2 in X}\sfil \\
\hline
Odd &\rlap{\hskip 1.2 in X}\sfil&\rlap{\hskip 1.2 in X}\sfil\\
\hline
\multicolumn{3}{|c|} {$\sigma=2$}\\ \hline
\head&Even&Odd\\
\hline
Even&\begin{tabular}{c}
           $K_{ss}(2)$ $K_{sv}^{(+)}(21)$ $K_{sv}^{(-)}(22)$ $K_{vv}(211)$
           \\
           $K_{vv}(202)$ $K_{vv}(220)$ $K_{vv}(222)$
     \end{tabular}& \rlap{\hskip 1.2 in X}\sfil\\
\hline
     Odd & \rlap{\hskip 1.2 in X}\sfil &
     $K_{sv}^{(+)}(22)$ $K_{sv}^{(-)}(21)$ $K_{vv}(212)$ $K_{vv}(221)$\\
\hline\hline
\end{tabular}}
\end{table}

\newpage
\def\stre{\phantom{\begin{tabular}{c}\makebox[0pt]{}\\ \end{tabular}}}
\begin{table}[h]
\caption{Expansion of the reduced matrix elements of the
non--relativistic hadronic one body weak current multipole
operators in terms of power series of the momentum transfer $\kappa=
|{\bf k}_e + {\bf k}_\nu|$.
Here $\tilde g_A = g_A\pm {\ov\Delta{2M}}g_T$.
\label{MultOpExpand}}
\vskip 0.2 in
\begin{tabular}{|c|cc|c|c|}
\hline\hline
 &Vector Current&&&Axial Vector Current\\
\hline
 $C_0$ &$\dsp F_1\left (A_1 - {\ov 1 6}{\kappa^2\over M^2} A_5\right)$ &
\stre&
$C_1^5$ &$\dsp-i{\ov\kappa {2M}}\left ({\ov 1\sqrt{3}}\left(g_A\mp g_T\right)
           A_2 + {\ov 2 3} g_A A_3\right )$\\
\hline
 $L_0$ & $\dsp{\ov\Delta\kappa}F_1\left (A_1 - {\ov 1
6}{\kappa^2\over M^2} A_5
    \right )$
&\stre& $L_1^5$ & $\dsp{\ov i\sqrt{3}}\tilde g_A
  \left ( A_2 - {\ov{\kappa^2} {6M^2}}A_7 + {\ov{\sqrt 2\kappa^2}{15 M^2}}A_8
  \right )$\\
\hline
$M_1$ & $\dsp{\ov{i\kappa}{2 M}} G_M \sqrt{\ov 2 3} A_2 - {\ov{i\kappa}{3M}}
          F_1A_4$ &
\stre & $E_1^5$ &
  $\dsp{\ov{i\sqrt{2}}\sqrt{3}}\tilde g_A
  \left ( A_2 - {\ov{\kappa^2}{6M^2}}A_7 - {\ov{\sqrt{2}\kappa^2}{30 M^2}}A_8
  \right )$\\
\hline
$C_2$ & $\dsp {\ov 1 {15}} {\ov{\kappa^2} M^2} F_1 A_6$ &
\stre & $M_2^5$ & $\dsp {\ov 1{15}} g_A {\ov{\kappa^2} M^2} A_9$\\
\hline
$L_2$ & $\dsp {\ov 1{15}}{\ov{\kappa\Delta} M^2} F_1 A_{6}$ &
\stre & $E_3^5$ & $\dsp {\ov {2i}{15\sqrt{7}}} g_A {\ov{\kappa^2} M^2}
A_{10}$\\
\hline
$E_2$ & $\dsp {\ov 1{15}}\sqrt{\ov 3 2}{\ov{\kappa\Delta} M^2} F_1A_6$ &
\stre &
$L_3^5$ & $\dsp{\ov i{15}}\sqrt{3\over 7} g_A {\ov{\kappa^2} M^2} A_{10}$
\\
\hline\hline
\end{tabular}
\hspace{0.3in}
\end{table}

\newpage
\begin{table}[h]
\caption{Definitions for the static multipole operators. Here
  $\braa{J_f}\ldots\kett{J_i}$ denotes the reduced matrix elements of
  the corresponding multipole operators. Vector $\sigma$ represents
the collection of three Pauli $2\times 2$ matrices. \label{StaticMop}}
\vskip 0.2 in
\begin{tabular}{|c|c|c|c|}
\hline\hline
$A_1$ &\hskip 15 mm $\dsp\sum_{i=1}^{A}\braa{J_f}\tau^{(\pm)}
Y_0\kett{J_i}$ \hskip 15mm&
$A_6$ &\hskip 15 mm $M^2\dsp\sum_{i=1}^{A}\braa{J_f}\tau^{(\pm)}
r^2Y_2\kett{J_i}$
\hskip 15 mm\\
\hline
$A_2$ & $\dsp\sum_{i=1}^A\braa{J_f}\tau^{(\pm)}
{\cal Y}_{10}\cdot\sigma\kett{J_i}$ &
$A_7$ & $M^2\dsp\sum_{i=1}^A\braa{J_f}\tau^{(\pm)}
r^2{\cal Y}_{10}\cdot\sigma\kett{J_i}
$\\
\hline
$A_3$ & $\dsp\sum_{i=1}^A\braa{J_f}\tau^{(\pm)}
rY_1\sigma\cdot\nabla\kett{J_i}$ &
$A_8$ & $M^2\dsp\sum_{i=1}^A \braa{J_f}\tau^{(\pm)}
r^2{\cal Y}_{12}\cdot\sigma\kett{J_i}
$\\
\hline
$A_4$ & $\dsp\sum_{i=1}^A\braa{J_f} \tau^{(\pm)}
r{\cal Y}_{11}\cdot\nabla\kett{J_i}$ &
$A_9$ & $M^2\dsp\sum_{i=1}^A\braa{J_f}\tau^{(\pm)}
r^2{\cal Y}_{22}\cdot\sigma\kett{J_i}
$\\
\hline
$A_5$ & $M^2\dsp\sum_{i=1}^A\braa{J_f} \tau^{(\pm)}r^2 Y_0\kett{J_i}$ &
$A_{10}$ & $M^2\dsp\sum_{i=1}^A\braa{J_f} \tau^{(\pm)}
r^2{\cal Y}_{32}\cdot\sigma
\kett{J_i}$\\
\hline\hline
\end{tabular}
\end{table}

\newpage
\begin{table}[h]
\caption{R--matrix parameters for the $J^P=0^+$ state obtained from
         a best fit to the experimental $\alpha$--$\alpha$ scattering
         phase shifts.\label{J0para}}
\vskip 0.2 in
\begin{tabular}{c|cccccc}
\hline
E $(MeV)$    &a $MeV^{-1}$& b $MeV^{-2}$&c $MeV^{-1}$&$w_0$ $(MeV)$ &
$r_0$ $(fm)$ & z $MeV^{-2}$  \\
\hline
  0.0    & 0.324  &--0.272 &--6.622 & 0.011  & 4.39 & 1.34$\times 10^{-4}$\\
\hline
\end{tabular}
\end{table}

\newpage
\begin{table}[h]
\caption{R--matrix parameters for the $J^P=2^+$ state obtained from a
         best fit to the experimental
         $\alpha$--$\alpha$ scattering
         phase shifts. No energy dependence for the doublet states
         located near $E\sim 16$ MeV is considered, which is
         indicated by $0$'s in the table. The $\alpha$--$\alpha$ phase
         shifts are very insensitive to the values of $w_0$ for the
         16 MeV doublet states in the energy region considered,
         they will be determined in the $\alpha$ particle
         spectra fit.  \label{J2para}}
\vskip 0.2 in
\begin{tabular}{c|cccccc}
\hline
E $MeV$    &a $MeV^{-1}$& b $MeV^{-2}$&c $MeV^{-1}$&$w_0$ (MeV) & $r_0$ (fm) &

z $MeV^{-2}$  \\
\hline
  3.36 & 0.145 &--0.01704 &  0.00216 & 1.80 &  & \\
 16.63 & 0     & 0       &  0       &  ?
  & 2.45  &5.2$\times 10^{-4}$\\
 16.92 & 0     & 0       &  0       &  ? &  &   \\
\hline
\end{tabular}
\end{table}

\newpage
\begin{table}[h]
\caption{R--matrix parameters for the $J^P=4^+$ state obtained from a best
         fit to the experimental
         $\alpha$--$\alpha$ scattering phase shifts.
         \label{J4para}}
\vskip 0.2 in
\begin{tabular}{c|cccccc}
\hline
E $MeV$    &a $MeV^{-1}$& b $MeV^{-2}$&c $MeV^{-1}$&$w_0$ (MeV)& $r_0$ (fm) &
z $MeV^{-2}$  \\
\hline
12.15      & --0.400 & --0.048 & --0.126 & 3.54 & 3.65 & 0.0 \\
\hline
\end{tabular}
\end{table}

\newpage
\begin{table}[h]
\caption{Values for various static single nucleon
weak current form factors used.
\label{FormFactors}}
\vskip 0.2 in
\begin{tabular}{cccc}
\hline
 $g_A$ & $F_1$ & $G_M^V$ & $g_T$ \cr
\hline
--1.254 & 1.00  & 4.71    & 0.00\cr
\hline
\end{tabular}
\end{table}

\newpage

\begin{table}[h]
\caption{Kinematic functions correspond to the contribution of the
symmetric (with respect to $A^{\mu\nu}$ and $B^{\mu\nu}$) and dispersive
parts of the final lepton--hadron Coulomb scattering T--matrix contribution.
\label{KineFuncCoul}}
\vskip 0.2 in
\footnotesize {
\begin{tabular}{|c|c|c|}\hline\hline
\multicolumn{3}{|c|} {$\sigma=2$}\\ \hline
\head&Odd&Even\\
\hline
Odd&\begin{tabular}{c}
           $K_{ss}(2)$ $K_{sv}^{(+)}(21)$ $K_{sv}^{(-)}(22)$ $K_{vv}(211)$
           \\
           $K_{vv}(202)$ $K_{vv}(220)$ $K_{vv}(222)$
     \end{tabular}& \rlap{\hskip 1.2 in X}\sfil\\
\hline
     Even & \rlap{\hskip 1.2 in X}\sfil &
     $K_{sv}^{(+)}(22)$ $K_{sv}^{(-)}(21)$ $K_{vv}(212)$ $K_{vv}(221)$\\
\hline\hline
\end{tabular}}
\end{table}

%
%  end tab
%
\end{document}